\documentclass[research]{fcs}
\usepackage{layout}
\usepackage{subfig}
\usepackage{hyperref}
\usepackage{enumitem}
\usepackage[dvipsnames]{xcolor}
\usepackage{algorithm}
\usepackage{algorithmic}

\title{iScheduler: Reinforcement Learning–Driven Continual Optimization for Large-Scale Resource Investment Problems}

\author[1,*]{Yi-Xiang~HU}
\author[1,*]{Yuke~WANG}
\author[1,+]{Feng~WU}
\author[1]{Zirui~HUANG}
\author[1]{Shuli~ZENG}
\author[1,+]{Xiang-Yang~LI}
\address[1]{School of Computer Science and Technology, University of Science and Technology of China, Hefei 230026, China}

\corremail{wufeng02@ustc.edu.cn, xiangyangli@ustc.edu.cn}
\fcssetup{
  received = {Jan 28, 2026},
  accepted = {Jul 15, 2026},
}

\begin{abstract}
Scheduling precedence-constrained tasks under shared renewable resources is critical to modern computing platforms. It is often modeled as the Resource Investment Problem (RIP) by minimizing the cost of provisioned renewable resources under precedence and timing constraints. Unfortunately, exact mixed-integer programming and constraint programming become impractically slow on large RIP instances, and dynamic updates require schedule revisions under tight latency budgets. To address this, we present iScheduler, a reinforcement-learning-driven iterative scheduling framework for large RIP. Specifically, it formulates RIP solving as a Markov decision process over decomposed subproblems and constructs schedules through sequential process selection. By doing this, the framework accelerates optimization and supports reconfiguration by reusing unchanged process schedules and rescheduling only affected processes. To evaluate this framework, we release L-RIPLIB, an industrial-scale benchmark derived from cloud-platform workloads with 1,000 instances of 2,500–10,000 tasks. Our experiments show that iScheduler attains competitive resource costs while reducing time to feasibility by up to 43$\times$ against leading solver-backed baselines. 
\end{abstract}
\keywords{Constraint Programming; Continual Optimization; Mixed-Integer Programming; Reinforcement Learning; Resource Investment Problem.}

\begin{document}

\section{Introduction}

Scheduling precedence-constrained tasks under limited shared resources (e.g., GPUs or network bandwidth) is fundamental to modern computing platforms~\cite{liu2024muxflow,Si2026CollaborativeRegistryPlanning}. 
However, in production settings, task parameters rarely remain static. For example, execution durations, feasible time windows, and resource requirements shift with workload demand, system congestion, and unexpected delays~\cite{CAI2024102628,10.1609/icaps.v33i1.27244,li2022scheduling}. Each fluctuation usually triggers a \textit{reconfiguration request} that must be handled under tight latency budgets. This makes fast schedule adaptation to dynamic task parameters essential but difficult.

In the literature, these scheduling settings are commonly formalized as Resource Investment Problems (RIPs), where the objective is to minimize the cost of allocating renewable resources while satisfying precedence and timing requirements~\cite{KRETER2018472}. RIPs are known to be NP-complete~\cite{10.5555/3491440.3491681}, and existing mainstream approaches often rely on Mixed Integer Programming (MIP) or Constraint Programming (CP). However, both MIP and CP solvers incur prohibitive runtimes on large instances. For example, a 10,000-task project with  millions of decision variables can require hours of computation even with a strong commercial solver (e.g., Gurobi~\cite{gurobi}).
This leads to our first challenge: (C1) \textit{How to generate high-quality schedules for large RIP instances within practical computation time limits?}

A commonly-known strategy for scaling to large instances is \textit{iterative decomposition}, which partitions the instance into smaller subproblems solved across multiple rounds~\cite{doi:10.1287/opre.1060.0358,LI1992370,LIU2021107553,copter}. However, we observe that the \textit{order}, in which tasks or task groups are scheduled, strongly influences the final solution quality (see \S \ref{Order}). Existing heuristic selection rules (e.g., critical-path, slack-based rules, or resource-profile heuristics) exhibit large performance variation across instances.
This leads to our second challenge: 
(C2) \textit{How to select scheduling orders that can outperform fixed heuristics and generalize across instances?}

Moreover, when task parameters change, re-solving the entire RIP from scratch is operationally unacceptable in practice. In many deployments, we observe that only a small local portion of the schedule is actually affected, yet traditional solvers require to recompute globally for the entire schedule.
Therefore, the core dynamic-setting challenge that we consider here is:
(C3) \textit{How to  update schedules efficiently under parameter variations without restarting the whole solving process?}

To address the aforementioned challenges, we propose iScheduler, a reinforcement-learning-driven iterative scheduling framework for large-scale RIPs. 
Specifically, our design separates the difficulty of large-scale continual scheduling into three coupled but distinct mechanisms. 
Firstly, we build a Directed Acyclic Graph (DAG) for the task and decompose the task DAG into process-level subproblems, reducing each solver invocation from a monolithic RIP to a much smaller exact local optimization problem. 
This decomposition primarily addresses the scalability challenge of C1. 
Secondly, because local subproblems are committed sequentially, their ordering determines the evolution of the global resource-pool usage and therefore affects the feasible region of later subproblems. 
We model this ordering problem as a Markov Decision Process (MDP) and learn a Reinforcement Learning (RL) based policy to select the next process under the current scheduling state, addressing the coordination challenge of C2. 
Finally, for each selected process, the local solver may produce multiple feasible schedules with different effects on future resource peaks. 
A ranking-based value network selects among these candidates before commitment, further aligning local decisions with the final global objective. 
Under reconfiguration, we reuse schedules of unaffected processes and reschedule only impacted ones, thereby addressing C3.

To summarize, our main contributions in this article are threefold:
\begin{itemize}[leftmargin=*, labelsep=0.5em]
\item \textbf{iScheduler Framework.} We first formulate large-scale RIP solving as a sequential decision process over decomposed subproblems. Then, we separately learn two policies: 1) a RL-based process-selection policy that orders which subproblem to commit next, and 2) a ranking-based solution-selection policy that picks among multiple feasible local schedules for each subproblem. Both of the policies condition on a shared Graph Neural Network (GNN) state representation of the process-level interaction graph and the evolving resource-pool usage. All together, they capture long-range interactions induced by shared resources and overlapping time windows. In addition, our framework supports reconfiguration by reusing unchanged process schedules and rescheduling only affected processes.
\item \textbf{L-RIPLIB Benchmark.} We release L-RIPLIB, an industrial-scale RIP dataset with 1,000 instances ranging from 2,500 to 10,000 tasks per instance. It is designed to evaluate large-scale optimization and reconfiguration settings beyond PSPLIB~\cite{kolisch1997psplib} for the community.
\item \textbf{Empirical Results.} On the L-RIPLIB benchmark, iScheduler matches competitive resource costs while reducing time to feasibility by up to 43$\times$ against strong solver-backed baselines. Under dynamic updates, it achieves lower reconfiguration latency and stronger solution quality than state-of-the-art baselines.
\end{itemize}

\section{Preliminaries and Problem Settings}\label{sec:preliminaries}

\subsection{Resource Investment Problem}\label{sec:rip}

We consider the {Resource Investment Problem} (RIP), a scheduling problem with precedence constraints, renewable resources, and time-window restrictions. Formally, an RIP instance is defined as a tuple:
$q=\langle T,\mathcal{R},P,r,c,d,$ $e,l\rangle$,
where $T$ is the set of non-preemptive tasks, $\mathcal{R}$ is the set of renewable resource types, and $P\subseteq T\times T$ is the set of precedence constraints. For each task $i\in T$, $d_i$ denotes its processing time, $e_i$ its earliest start time, and $l_i$ its latest completion bound. Each task consumes $r_{i,k}$ units of resource $k\in\mathcal{R}$ while it is active. Provisioning one unit of resource $k$ incurs cost $c_k$, and $R_k$ denotes the project-wide provisioned capacity of resource $k$.

A schedule assigns each task $i\in T$ a start time $S_i$. Let $\mathcal{U}(S,t)$ denote the set of tasks active at time $t$ under schedule $S$. The RIP seeks a feasible schedule together with resource capacities that minimize the total provisioning cost:

\begin{equation}
    \min \;\;  \sum_{k \in \mathcal{R}} c_k R_k \label{eq1} \end{equation}\begin{equation}
    \text{s.t.}\;\;
     S_i + d_i \le S_j, \qquad \forall (i,j)\in P, \label{eq2} \end{equation}\begin{equation}
     \sum_{i \in \mathcal{U}(S,t)} r_{i,k} \le R_k, \qquad \forall k\in\mathcal{R},\ \forall t, \label{eq3} \end{equation}\begin{equation}
     e_i \le S_i \le l_i-d_i, \qquad \forall i\in T, \label{eq4} \end{equation}\begin{equation}
     R_k \ge 0, \qquad \forall k\in\mathcal{R}. \label{eq5}
\end{equation}
Constraint~\eqref{eq2} enforces precedence feasibility, Constraint~\eqref{eq3} limits the instantaneous usage of each renewable resource, and Constraint~\eqref{eq4} enforces task-specific time windows. The objective in~\eqref{eq1} minimizes the total cost of renewable resource investment.

\subsection{Continual Reconfiguration Setting}\label{sec:reconfiguration}

In practical deployments, RIP instances evolve over time. Task durations, resource demands, feasible windows, or even precedence relations may change after an initial schedule has already been produced and partially deployed. Let
$q=\langle T,\mathcal{R},P,r,c,d,e,l\rangle$
be the original instance with an existing feasible schedule $(S_i)_{i\in T}$, and let
$q'=\langle T',\mathcal{R},P',r',c,d',e',l'\rangle$
be the updated instance after a reconfiguration request.

A \emph{classical} optimizer solves the updated instance from scratch,
$h(q') \rightarrow (S'_i)_{i\in T'},$
without using the previously committed schedule. In contrast, a \emph{continual reconfiguration} method exploits prior decisions and updates only the affected part of the solution,
$f\bigl(q,(S_i)_{i\in T},q'\bigr) \rightarrow (\widetilde{S}_i)_{i\in T'}$.
The goal is to preserve feasibility and solution quality while reducing recomputation latency. This setting is especially important in large-scale systems, where dynamic updates are frequent but full re-optimization is too slow for operational use.
\subsection{Process-Level Abstraction}
\label{sec:process_abstraction}

To support scalable iterative optimization, we introduce a process-level
abstraction over the task graph. Given the task-level directed acyclic
graph \(G_T=(T,P)\), we obtain processes by taking the weakly connected
components of \(G_T\) after ignoring edge directions. Let
\(\{\mathcal{P}_1,\mathcal{P}_2,\ldots,\mathcal{P}_n\}\) denote the resulting
processes, and let \(T_{\mathcal{P}_i}\subseteq T\) be the set of tasks contained
in process \(\mathcal{P}_i\). These task subsets form a partition of \(T\) and
serve as the basic scheduling units of iScheduler.

This abstraction targets operational workloads composed of multiple job
chains, service groups, or request groups that share renewable resources
while remaining separated by precedence structure. Under weak-component
decomposition, every precedence edge belongs to exactly one process; hence
no cross-process precedence constraint is removed or relaxed. The decomposition
therefore preserves precedence feasibility by construction. The remaining
interaction among processes arises from shared resource usage over time rather
than from precedence constraints. When the task DAG forms a single weak
component, the decomposition degenerates to one process and the ordering policy
has no process-level decision to make. Handling such cases would require an
additional graph-partitioning layer with explicit boundary-constraint handling,
which is outside the scope of this work.

Based on the process partition, we construct a process-level interaction graph
\(G_{\mathrm{PL}}=(V_{\mathrm{PL}},E_{\mathrm{PL}})\). Each node in
\(V_{\mathrm{PL}}\) corresponds to one process. An undirected edge
\((\mathcal{P}_i,\mathcal{P}_j)\in E_{\mathrm{PL}}\) is added when the two
processes have potential temporal overlap and thus may compete for shared
renewable resources. Specifically, an edge exists if there are tasks
\(u\in T_{\mathcal{P}_i}\) and \(v\in T_{\mathcal{P}_j}\) whose activity
envelopes overlap:
$[e_u,l_u]\cap [e_v,l_v]\neq \emptyset$.
This edge does not imply that the two processes must conflict in the final
schedule; rather, it records the possibility of simultaneous resource demand.
The actual contention level depends on task demands, resource costs, committed
schedules, and the evolving resource-usage profile.

We use the Resource Pool Usage (RPU) profile to summarize this evolving
global context. For each resource \(k\in \mathcal{R}\) and time \(t\), let
\(u_k(t)\) denote the amount of resource \(k\) already occupied by committed
processes. The full usage profile is
$\mathrm{RPU}=\{u_k(t)\}_{k\in \mathcal{R}}$.
RPU is passed from one scheduling iteration to the next. Once a process is
committed, its resource demand is added to RPU, thereby changing the residual
capacity available to unscheduled processes. Thus, \(G_{\mathrm{PL}}\) captures
the static possibility of inter-process contention, while RPU captures the
dynamic effect of previous commitments.

\subsection{Iterative Optimization Interface}\label{sec:iterative_interface}

The continual optimization problem is solved through an iterative interface defined over the process-level abstraction. Under reconfiguration, processes unaffected by the update retain their previously committed schedules, while affected processes are marked as unscheduled. Let $\mathcal{C}_{PL}\subseteq V_{PL}$ denote the current set of unscheduled processes.

At iteration $v$, the solver selects one process $\mathcal{P}_v\in\mathcal{C}_{PL}$ and constructs a local subproblem, denoted by $\mathrm{RIP}_v$, over the task subset $T_{\mathcal{P}_v}$. This subproblem is conditioned on the current RPU, so previously committed schedules remain fixed and only the selected process is optimized. The solver then generates one or more feasible local schedules for $\mathcal{P}_v$, after which one local schedule is committed. Once committed, the corresponding task start times are fixed, the RPU is updated, and $\mathcal{P}_v$ is removed from $\mathcal{C}_{PL}$.

This procedure repeats until no unscheduled processes remain. The final schedule is obtained by combining the retained schedules of unchanged processes with the newly committed schedules of updated processes. Based on this iterative interface, \S \ref{sec:iScheduler} presents iScheduler, which learns how to choose the next process and how to commit local solutions effectively under dynamic reconfiguration.

\section{Related Work}\label{sec:related_work}

\subsection{Resource Investment Problem and Exact Optimization}

The RIP and closely related resource-availability-cost scheduling models have been studied primarily through exact optimization frameworks. Classical formulations rely on MIP and CP to jointly optimize task start times and renewable resource capacities under precedence and time-window constraints \cite{KRETER2018472,10.5555/3491440.3491681}. These formulations provide a rigorous treatment of feasibility and objective optimality, and they remain the dominant choice for small and medium-sized instances where strong bounds and exact solutions are required.

Despite their modeling power, exact RIP solvers face severe scalability limits on large instances. As the number of tasks, precedence arcs, and time-indexed resource interactions grows, both MIP and CP formulations experience a sharp increase in decision variables, propagation complexity, and search depth \cite{KRETER2018472,10.5555/3491440.3491681}. This challenge becomes more pronounced in practical settings where schedules must be updated repeatedly after parameter changes. In such cases, solving each updated instance from scratch often incurs prohibitive latency even when high-quality commercial solvers are available.

\subsection{Decomposition for Large-Scale Scheduling}

A standard strategy for scaling difficult scheduling problems is decomposition. Prior work has explored decomposition-based genetic algorithms, iterative scheduling procedures, and multi-stage decomposition schemes that partition a large problem into smaller components and solve them progressively \cite{doi:10.1287/opre.1060.0358,LI1992370,LIU2021107553}. These methods reduce the size of each optimization step and therefore improve tractability on large problem instances.

Decomposition, however, does not eliminate the core coordination difficulty. Once a problem is split into multiple parts, the final solution quality depends not only on the quality of each local solve but also on how subproblems are defined, in what order they are processed, and how intermediate decisions constrain the remaining search space. In scheduling problems with shared renewable resources and coupled time windows, local commitments can alter future feasibility and resource peaks in ways that are difficult to anticipate using static rules. Consequently, decomposition alone improves computational tractability, but it does not fully resolve the decision problem induced by sequential local commitment.

A closely related system-level example is POP, which accelerates large-scale resource allocation by partitioning the global problem and solving subproblems in parallel \cite{narayanan2021solving}. POP demonstrates that partition-based optimization can yield substantial speedups when cross-partition coupling is sufficiently weak. 

\subsection{Continual Optimization and Schedule Reconfiguration}

Many real scheduling systems operate in dynamic environments, where durations, release times, deadlines, or resource requirements change after an initial schedule has already been produced. This has motivated research on schedule repair, incremental re-optimization, warm-started solving, and disruption-aware rescheduling. In the broader scheduling literature, these methods seek to preserve useful parts of an existing schedule while adapting the affected regions efficiently, rather than discarding all prior decisions and solving from scratch.

Among the works most relevant to our setting, COpter formulates large-scale resource allocation as a continual optimization problem and accelerates re-optimization by reusing previous solutions as warm starts \cite{copter}. This line of work highlights an important practical principle: under dynamic updates, reusing prior commitments is often essential for meeting latency budgets. However, warm-start-based continual optimization still leaves open a key question for large, structured scheduling instances: when only part of the instance must be rescheduled, which part should be optimized next, and how should local alternatives be committed so that later decisions remain favorable.

\subsection{Learning-Guided Scheduling and Combinatorial Optimization}

Recent years have seen growing interest in learning-guided methods for combinatorial optimization and scheduling~\cite{hu2025ffcg}. In resource-constrained scheduling, deep reinforcement learning  has been used to handle dynamic disturbances and disruptions, while GNNs have been employed to encode precedence structure and resource interactions for fast decision making \cite{CAI2024102628,10.1609/icaps.v33i1.27244}. More broadly, learning-based methods have shown that structured neural representations can improve dispatching, ranking, and search guidance in discrete optimization problems.

These approaches demonstrate two important insights. First, scheduling quality often depends on sequential decisions whose long-term effects are difficult to capture using fixed priority rules. Second, graph-based encoders are well suited to scheduling domains because the underlying state naturally forms task or interaction graphs. At the same time, many learning-based schedulers optimize through direct policy construction or heuristic dispatching, and therefore do not preserve exact local feasibility in the way solver-backed optimization does. This trade-off becomes important in industrial RIP settings, where hard constraints and solution validity remain non-negotiable.

\section{iScheduler Framework}\label{sec:iScheduler}

\begin{figure*}[!tb]
    \centering
    \includegraphics[width=0.98\linewidth]{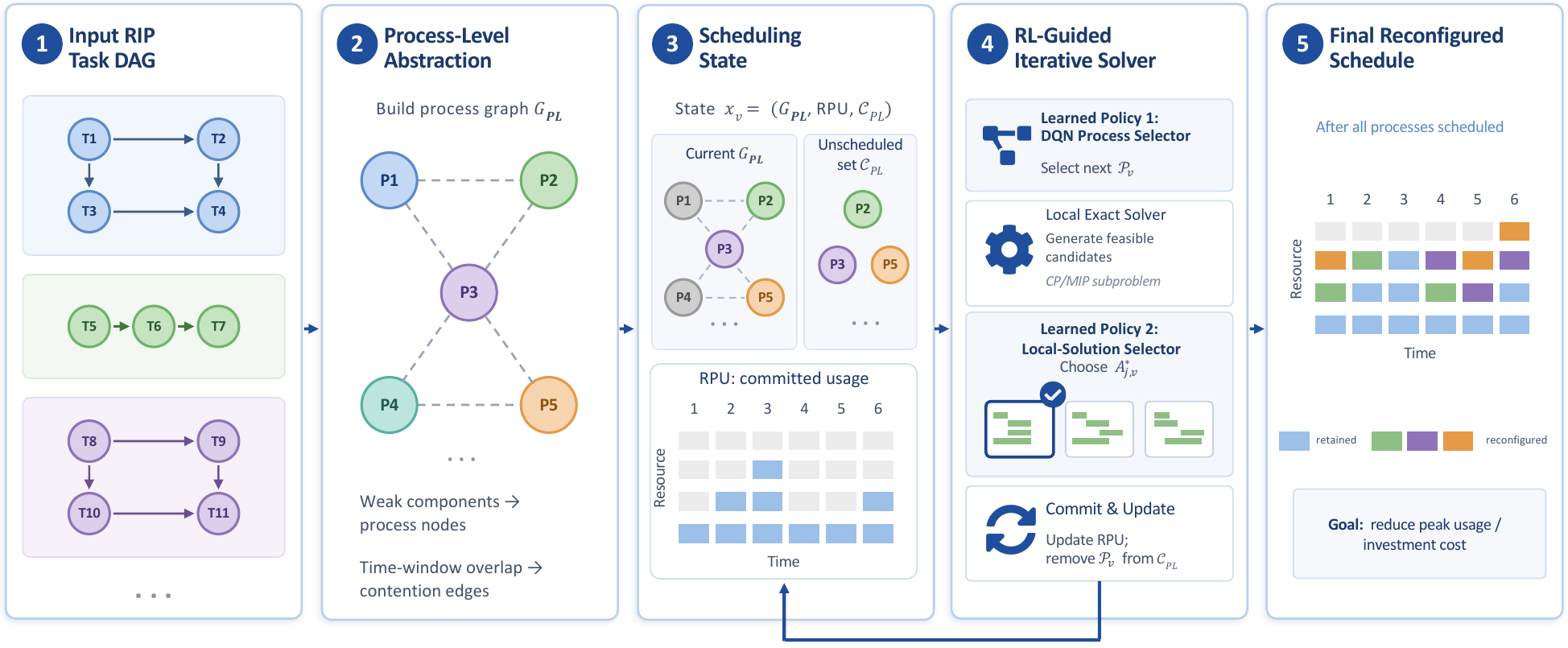}
    \caption{Overview of iScheduler. 
The input RIP is represented as a task-level DAG and abstracted into process-level scheduling units. 
A process-level interaction graph captures potential resource contention through feasible-window overlap, while the resource-pool usage (RPU) records the footprint of already committed schedules. 
At each iteration, the DQN policy selects one unscheduled process, an exact local solver generates feasible candidate schedules, and a learned selector commits one candidate. 
The committed schedule updates the RPU and the unscheduled process set. 
Under reconfiguration, unaffected processes retain their previous schedules, and only affected processes are rescheduled.}
\label{framework}
\end{figure*}

\begin{figure}[!t]
    \centering
    \includegraphics[width=0.45\textwidth]{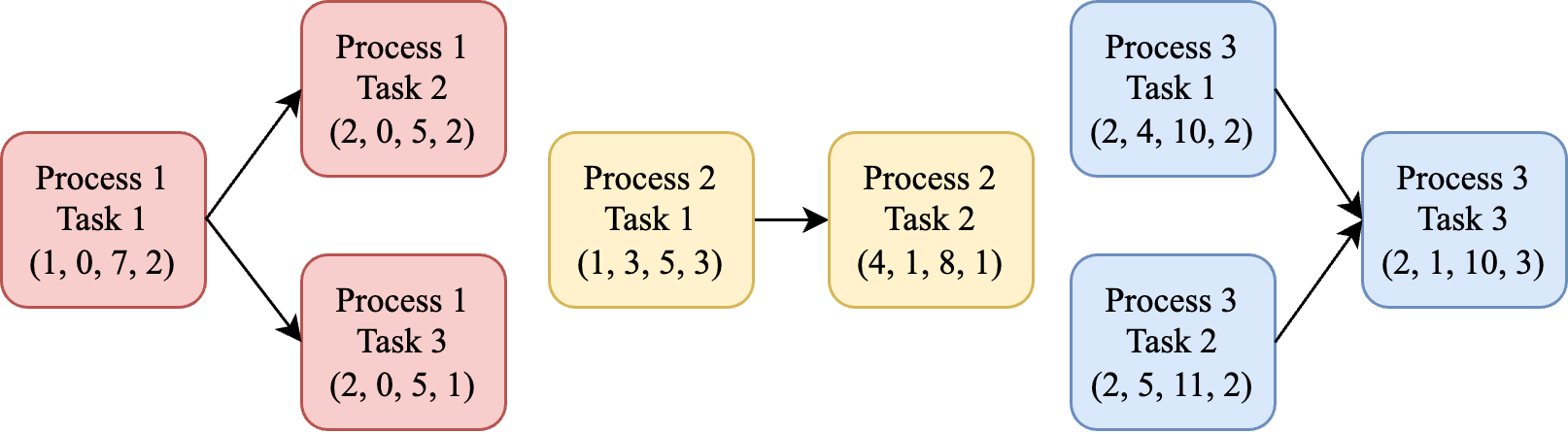}
    \caption{Task structures of three processes in an RIP. Each block represents a task, annotated as \((d_i, e_i, l_i, r_{i,1})\), where \(d_i\) is the duration, \(e_i\) the earliest start time, \(l_i\) the deadline, and \(r_{i,1}\) the demand for resource~1. Directed arrows denote precedence constraints: a task must finish before any successor can start.}
    \label{example_1}
\end{figure}

Figure~\ref{framework} gives a high-level overview of iScheduler.
The framework first maps the task-level RIP instance to a process-level abstraction, then repeatedly selects one unscheduled process, solves its local subproblem, commits one feasible local schedule, and updates the global scheduling state.
As a running example, Figure~\ref{example_1} shows three processes with their task structures, which we use throughout this section to illustrate scheduling decisions.
The following subsections describe the local subproblem, state representation, process-selection policy, candidate generation, local-solution selector, and reconfiguration procedure.

\subsection{Local Subproblem and Iterative Capacity Accounting}\label{sec:subproblem}

\begin{figure*}[!t]
\centering
\subfloat[Case 1, scheduling $\mathcal{P}_1$]{\includegraphics[width=0.3\linewidth]{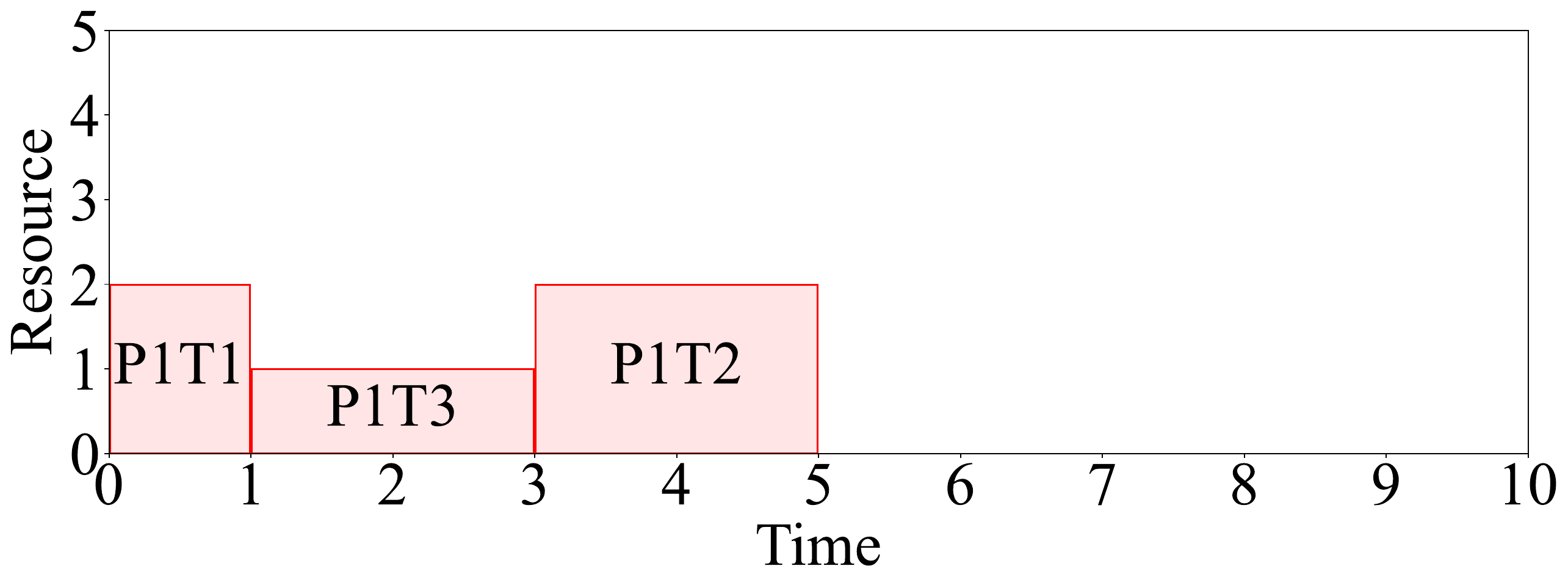}%
\label{fig_first_case}}
\hfil
\subfloat[Case 1, scheduling $\mathcal{P}_2$]{\includegraphics[width=0.3\linewidth]{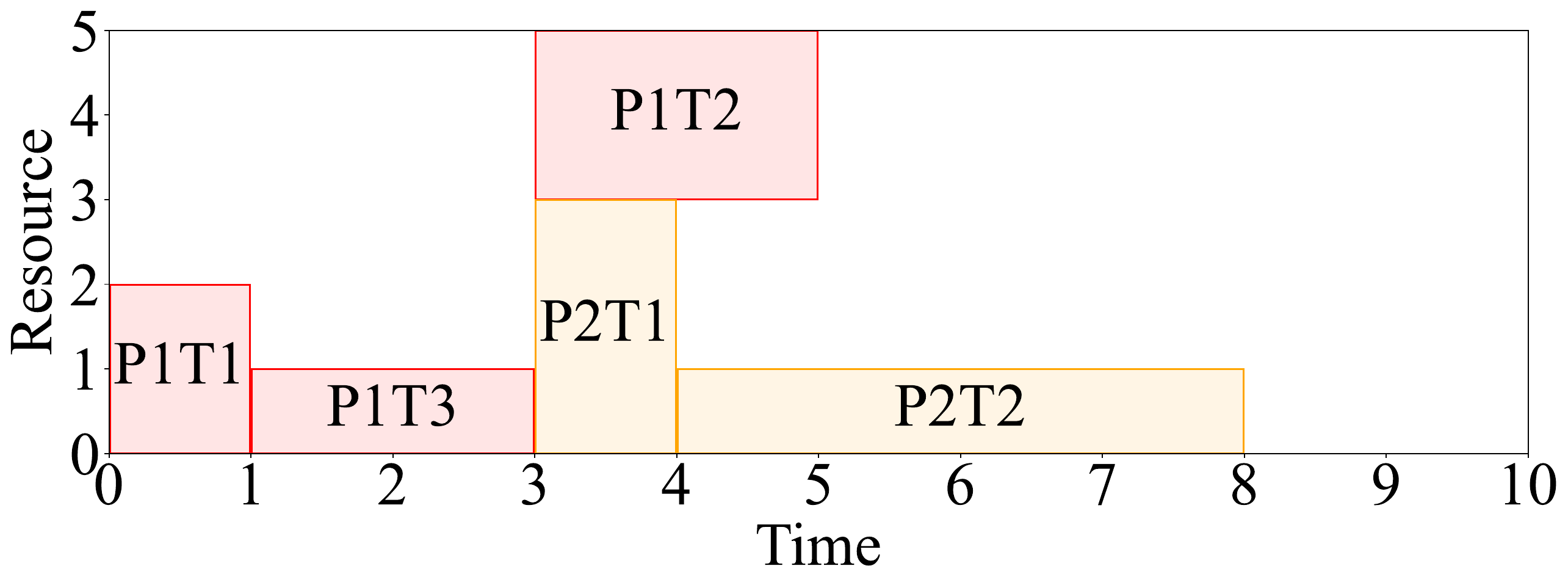}%
\label{fig_second_case}}
\hfil
\subfloat[Case 1, scheduling $\mathcal{P}_3$]{\includegraphics[width=0.3\linewidth]{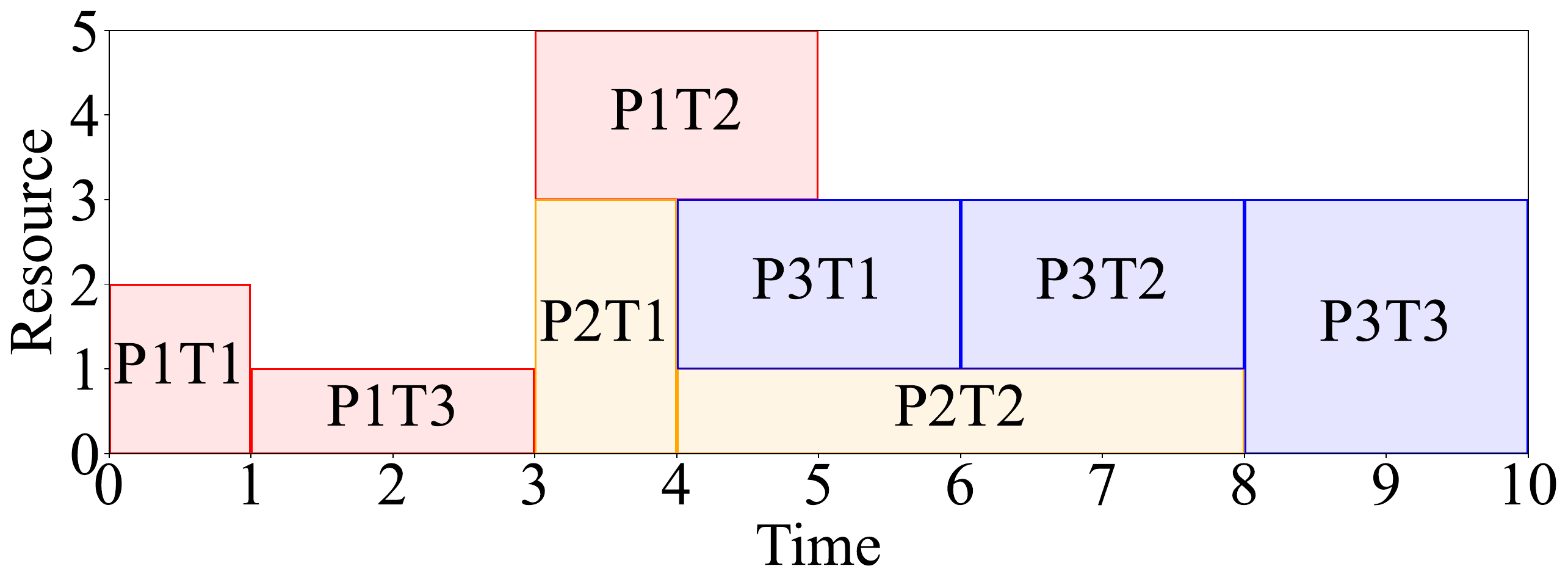}%
\label{fig_second_case2}}\\
\subfloat[Case 2, scheduling $\mathcal{P}_1$]{\includegraphics[width=0.3\linewidth]{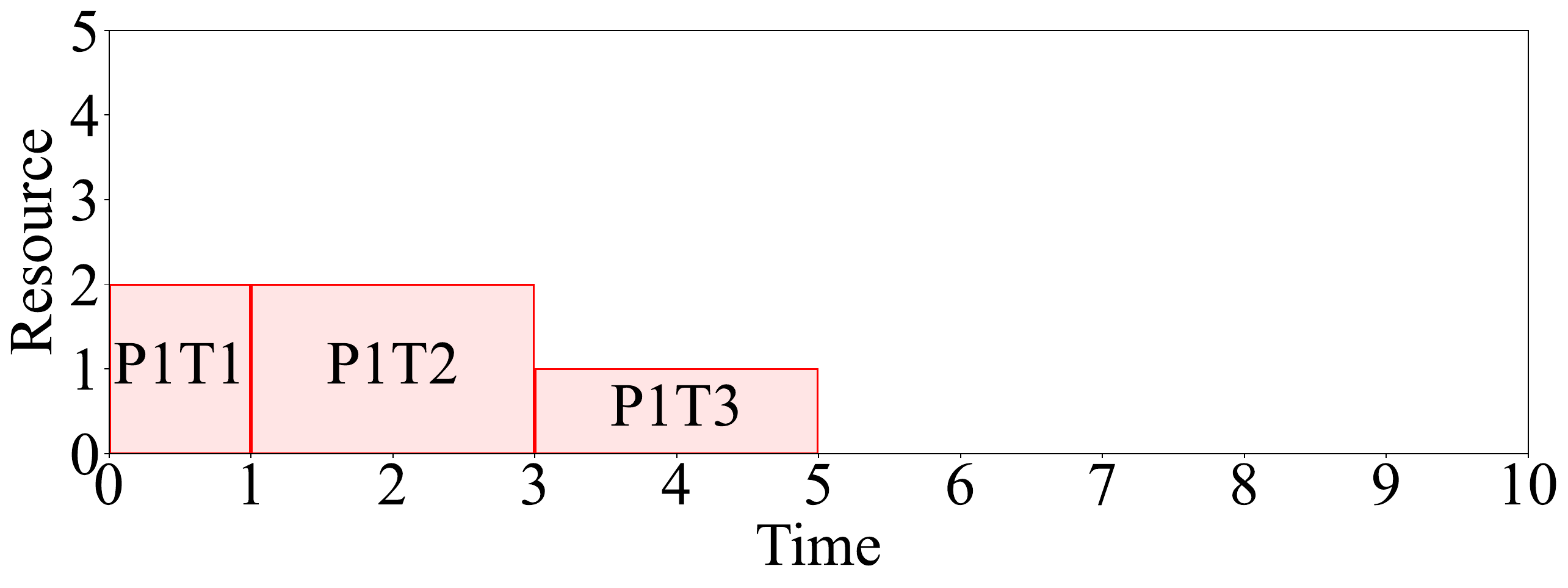}%
\label{fig_first_case1}}
\hfil
\subfloat[Case 2, scheduling $\mathcal{P}_2$]{\includegraphics[width=0.3\linewidth]{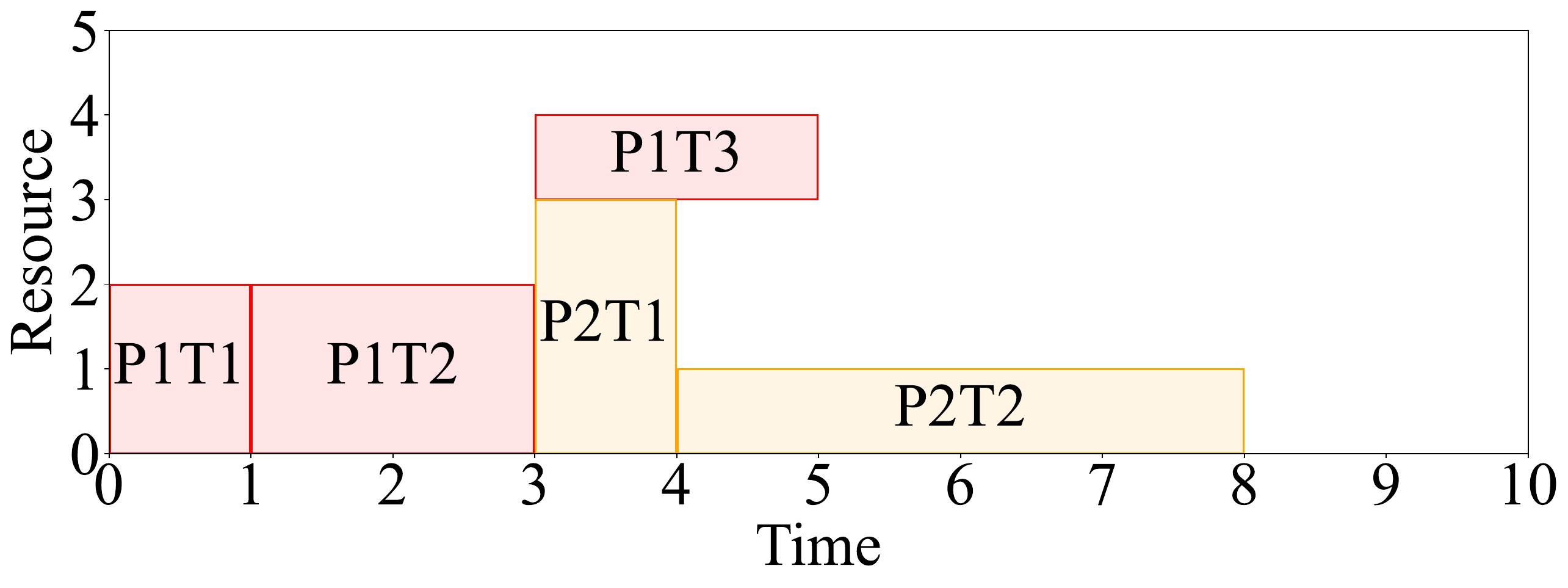}%
\label{fig_second_case1}}
\hfil
\subfloat[Case 2, scheduling $\mathcal{P}_3$]{\includegraphics[width=0.3\linewidth]{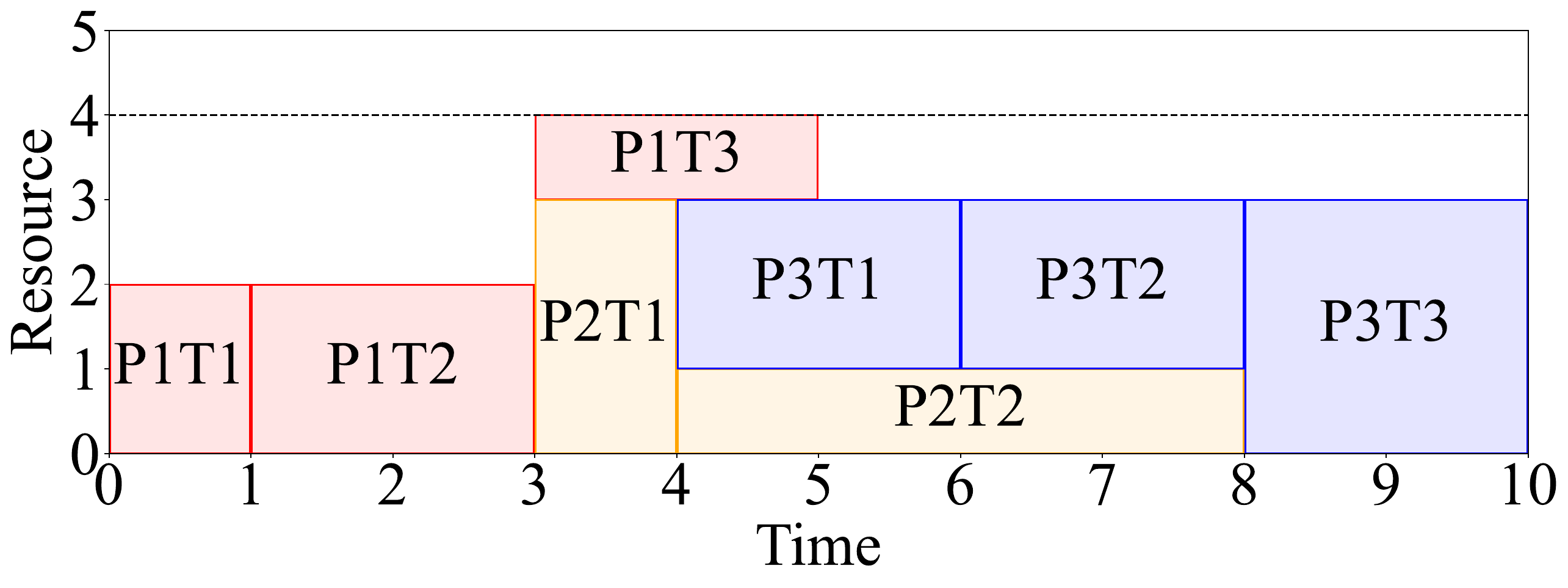}%
\label{fig_second_case21}}\\
\subfloat[Case 3, scheduling $\mathcal{P}_2$]{\includegraphics[width=0.3\linewidth]{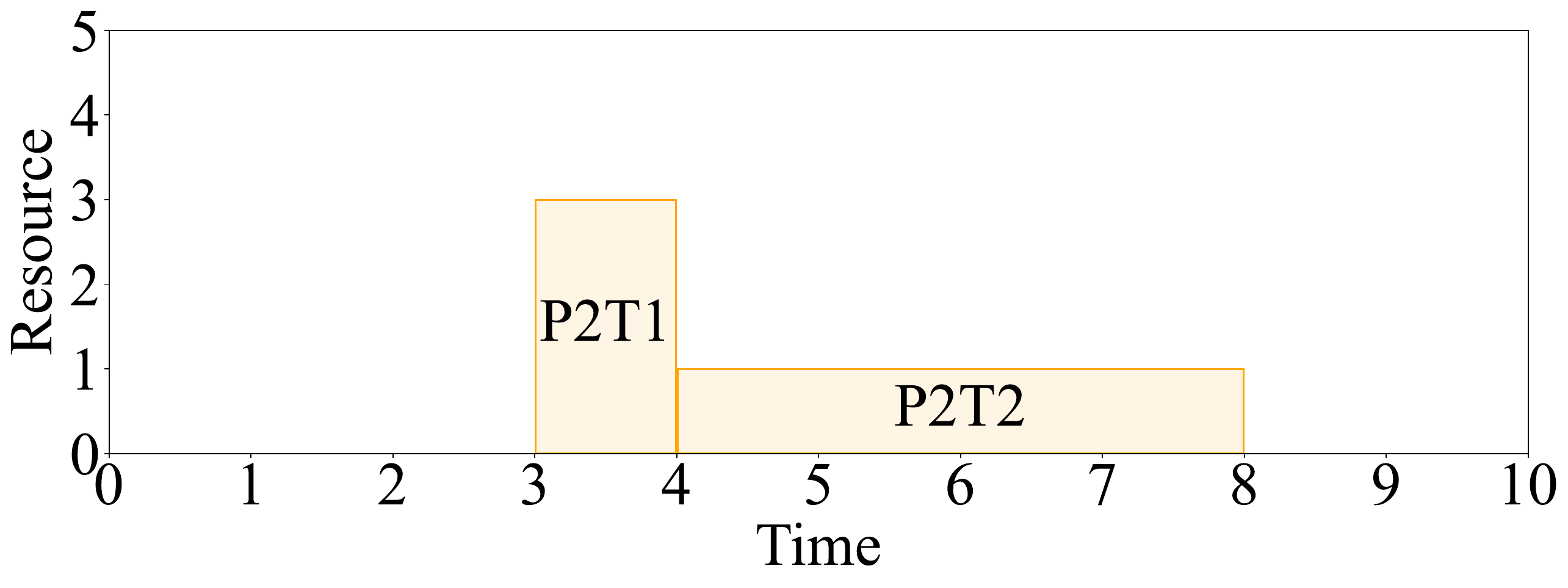}%
\label{fig_first_case3}}
\hfil
\subfloat[Case 3, scheduling $\mathcal{P}_3$]{\includegraphics[width=0.3\linewidth]{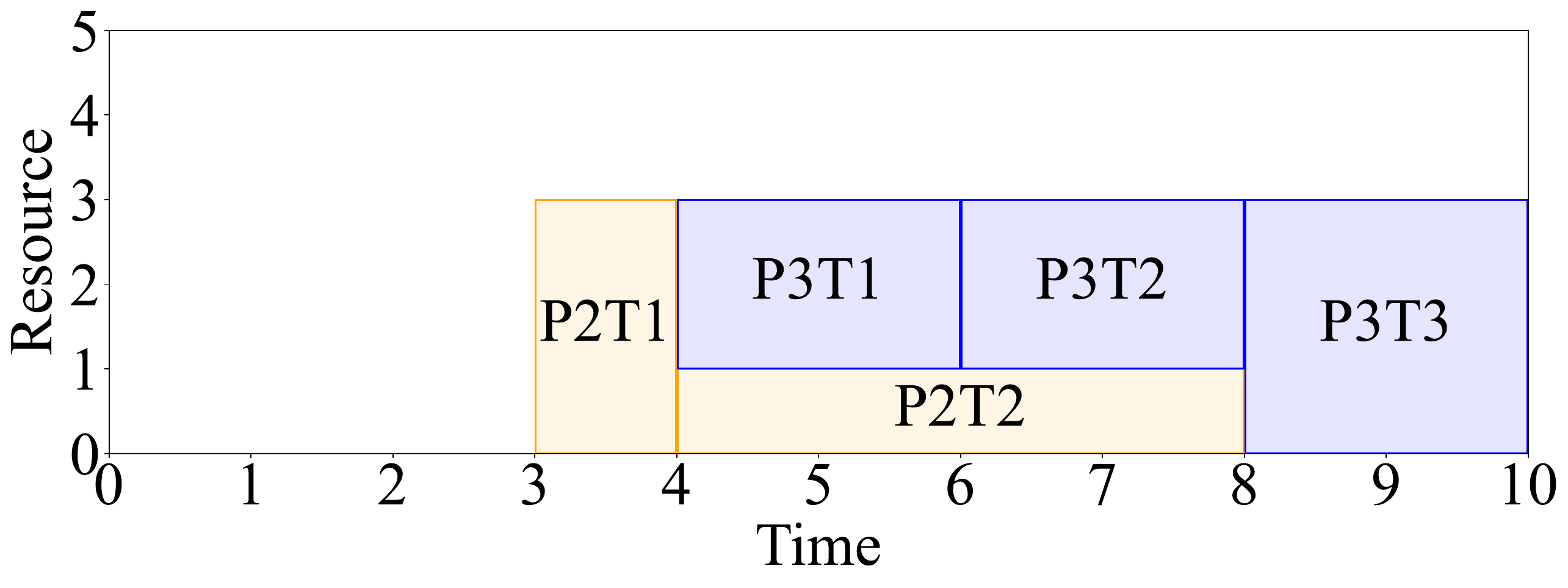}%
\label{fig_second_case3}}
\hfil
\subfloat[Case 3, scheduling $\mathcal{P}_1$]{\includegraphics[width=0.3\linewidth]{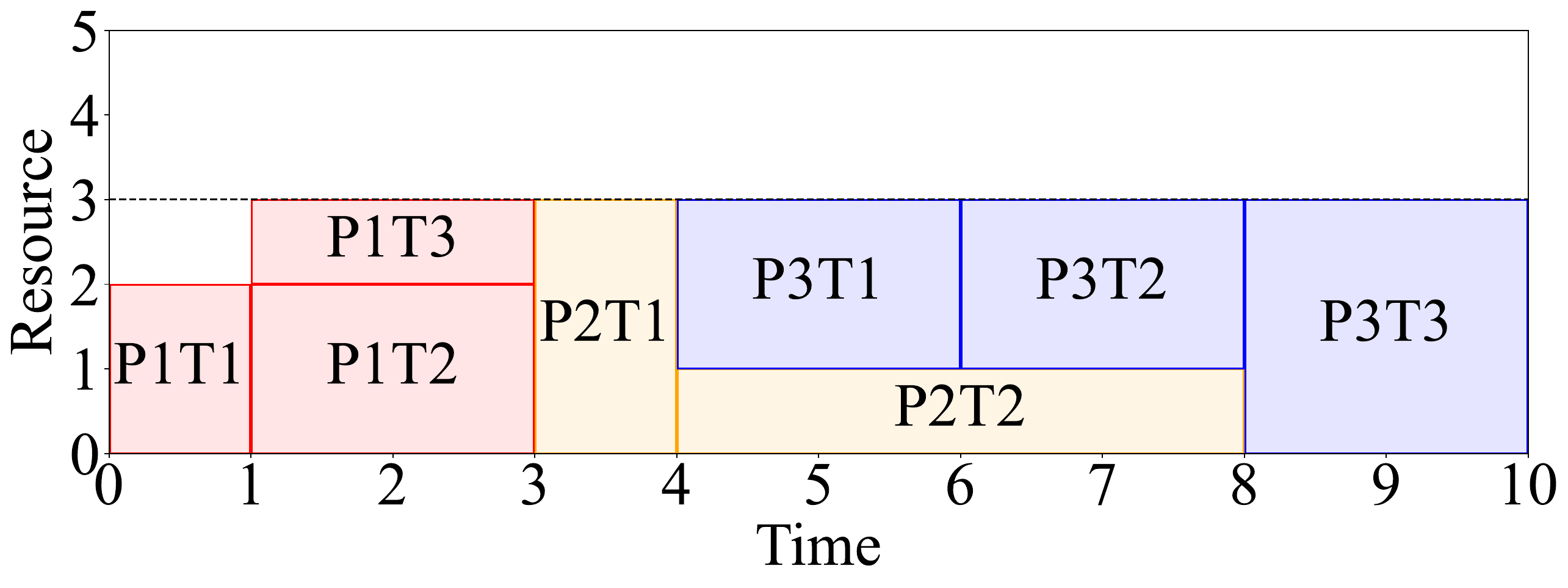}%
\label{fig_second_case32}}
    \caption{Effect of scheduling order and local solution selection on final performance. 
Each $\mathcal{P}_i$ denotes a process, and each task in a process is labeled as $\mathrm{P}_i\mathrm{T}_j$, where $i$ indexes the process and $j$ indexes the task within that process.
Different scheduling orders and local schedule choices lead to different resource usage outcomes. 
Case~1: Scheduling $\mathcal{P}_1 \rightarrow \mathcal{P}_2 \rightarrow \mathcal{P}_3$ results in total resource usage $R_1 = 5$. 
Case~2: Using the \emph{same} scheduling order but selecting an alternative local solution for $\mathcal{P}_2$ yields $R_1 = 4$. 
Case~3: Changing the scheduling order to $\mathcal{P}_2 \rightarrow \mathcal{P}_3 \rightarrow \mathcal{P}_1$ further reduces usage to $R_1 = 3$.}

\label{fig_sim}
\end{figure*}

At iteration \(v\), the selected process $\mathcal{P}_v$ is scheduled under the current RPU $\{u_k(t)\}_{k\in\mathcal{R}}$, with previously committed tasks fixed. The resulting local subproblem $\mathrm{RIP}_v$ retains the RIP structure of Eqs.~\eqref{eq1}--\eqref{eq5}, but is restricted to $T_{\mathcal{P}_v}$ and uses the residual budget $R_k^{(v)}-u_k(t)$:
% \begin{align}
\begin{equation}
    \min \;\;  \sum_{k \in \mathcal{R}} c_k \, R_k^{(v)} \end{equation}\begin{equation}
    \text{s.t.}\;\;
     S_i + d_i \le S_j \qquad \forall (i,j) \in P,\; i,j \in T_{\mathcal{P}_v}, \end{equation}\begin{equation}
     \sum_{i \in \mathcal{U}(S,t) \cap T_{\mathcal{P}_v}} r_{i,k}
      \;\le\; R_k^{(v)} - u_k(t)
      \qquad \forall k \in \mathcal{R},\; \forall t, \end{equation}\begin{equation}
     e_i \le S_i \le l_i-d_i \qquad \forall i \in T_{\mathcal{P}_v}, \end{equation}\begin{equation}
     R_k^{(v)} \ge 0 \qquad \forall k \in \mathcal{R}.
\end{equation}

For a representative $|T|=10{,}000$ instance, the monolithic formulation contains 3,706,353 variables and 43,701 constraints, whereas decomposition yields 330 processes whose subproblems contain 3,748.5 variables and 1,482.7 constraints on average, substantially reducing the burden on the exact solver at each iteration.

\textbf{Iterative accounting of capacity and usage.}
We maintain two quantities during the iterative procedure: the current usage profile $u_k^{(v)}(t)$ induced by previously committed processes, and the instantaneous demand of the newly committed process,
\begin{equation}
d_k^{(v)}(t)=\sum_{i\in \mathcal{U}(S,t)\cap T_{\mathcal{P}_v}} r_{i,k}.
\end{equation}
The resource constraints are equivalently $u_k^{(v)}(t)+d_k^{(v)}(t)\le R_k^{(v)}$ for all $k$ and $t$. After committing $\mathcal{P}_v$, we update the usage profile additively,
\begin{equation}
u_k^{(v+1)}(t)\;\leftarrow\;u_k^{(v)}(t)+d_k^{(v)}(t),
\end{equation}
and set the corresponding provisioned capacity by
\begin{equation}
R_k^{(v+1)}\;\leftarrow\;\max_{t}\,u_k^{(v+1)}(t).
\label{eq:Rk_update}
\end{equation}
After $V$ iterations, the final capacity is $R_k=R_k^{(V)}$ and the final objective is $\sum_{k\in \mathcal{R}} c_k R_k^{(V)}$. Each task contributes to $u_k$ only when its own process is committed, so the update is exact and introduces neither double-counting nor under-provisioning.

\subsection{State Representation}\label{Feature}

The state combines the process-level interaction graph $G_{PL}$ with the current RPU and is shared by both the process-selection policy and the local-solution selector. Each process node carries four features: intrinsic processing time $PT_{\mathcal{P}}=\sum_{i \in T_{\mathcal{P}}} d_i$, average weighted resource demand
\[
    \mathrm{WRD}_{\mathcal{P}}
    =
    \frac{1}{|T_{\mathcal{P}}| \sum_{k \in \mathcal{R}} c_k}
    \sum_{i \in T_{\mathcal{P}}} \sum_{k \in \mathcal{R}} c_k r_{i,k},
\]
the scheduling-order encoding (0 if unscheduled, otherwise the commit iteration), and the weighted resource usage in its feasible window
\[
    \mathrm{WRU}_{\mathcal{P}}(ES_{\mathcal{P}},LF_{\mathcal{P}})
    =
    \frac{1}{\sum_{k \in \mathcal{R}} c_k}
    \sum_{k \in \mathcal{R}}
    \frac{c_k}{R_k}
    \sum_{t = ES_{\mathcal{P}}}^{LF_{\mathcal{P}}} u_k(t),
\]
where $ES_{\mathcal{P}} = \min_{i\in T_{\mathcal{P}}} e_i$ and $LF_{\mathcal{P}} = \max_{i\in T_{\mathcal{P}}}l_i$. Each edge $(\mathcal{P}_i,\mathcal{P}_j)\in G_{PL}$ carries five features: the overlap window $(t_\mathrm{start}, t_\mathrm{end})$, the ratio of shared resource types $\frac{|\mathcal{R}_{\mathcal{P}_i} \cap \mathcal{R}_{\mathcal{P}_j}|}{|\mathcal{R}|}$, the two per-process overlap-window usages $\mathrm{WRU}_{\mathcal{P}_i}(t_\mathrm{start},t_\mathrm{end})$ and $\mathrm{WRU}_{\mathcal{P}_j}(t_\mathrm{start},t_\mathrm{end})$, their combined weighted utilization, and the normalized system-wide utilization in the same window.

We encode this graph with a GATv2 network~\cite{brody2021attentive}. 
Although node and edge annotations are defined locally, the resulting state representation is not purely local: the RPU is a global usage profile, and stacked message passing propagates contention information beyond immediate neighbors. 
This representation does not guarantee globally optimal ordering, but it provides the policy with the structural and resource-usage context needed to distinguish process orders that lead to different peak-resource trajectories.

\subsection{Process-Selection Policy}\label{Order}

\textbf{Why scheduling order matters.}
Decomposition reduces each solver call but leaves one global degree of freedom: the order in which processes are committed. Different orderings induce different RPU trajectories, which alter the feasible region seen by later subproblems. Figure~\ref{fig_sim} shows this directly: even under the same decomposition, changing either the process order or the committed local solution changes the final resource peak. This sequential dependence motivates learning a context-dependent ordering policy rather than relying on a fixed priority rule.

\begin{figure}[!t]
    \centering
    {\includegraphics[width=0.47\textwidth]{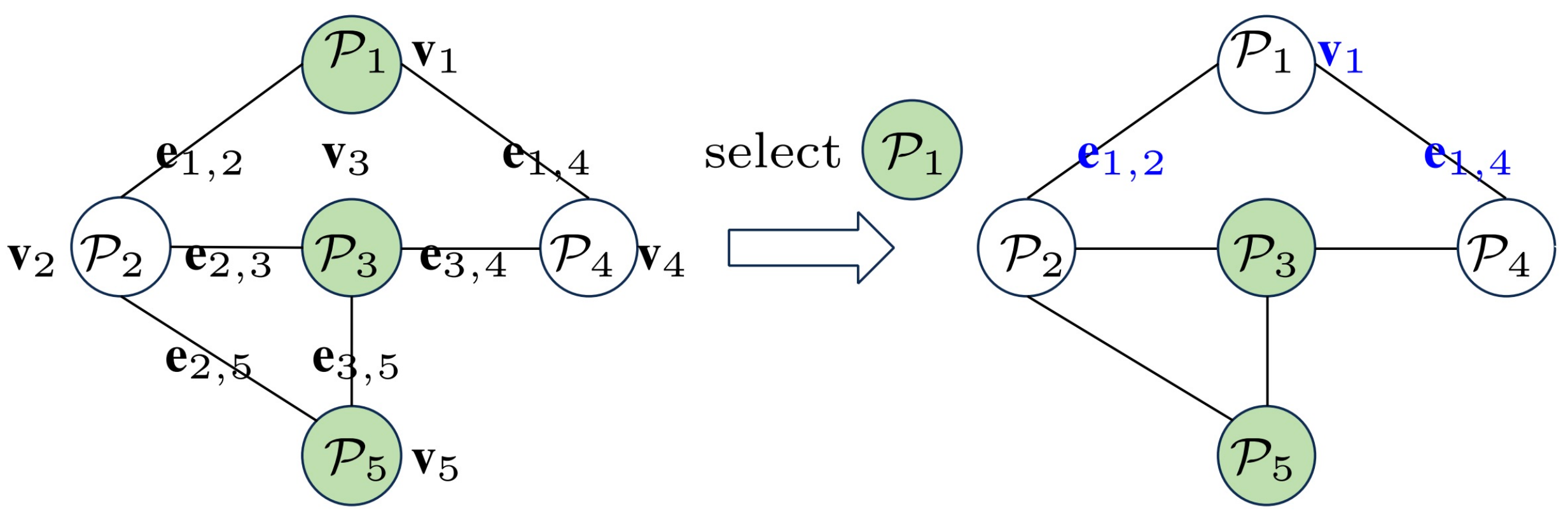}}
    \caption{State transition: when one of the three considered action nodes (in green) is selected, it transitions to a new state and updates the associated features (in blue).}
    \label{fig:state}
\end{figure}

\textbf{MDP formulation.}
We model the iterative procedure as an MDP $\langle \mathcal{X}, \mathcal{A}, \mathcal{T}, \mathcal{W}, \gamma\rangle$. The state $\mathbf{x}$ is the annotated process graph together with the current RPU; the action set is the unscheduled process set $\mathcal{A}(\mathbf{x})=\mathcal{C}_{PL}$; and transitions are deterministic because selecting $\mathcal{P}_v$ uniquely determines subproblem construction, local commitment, and the next state $\mathbf{x}_{v+1}$; Figure~\ref{fig:state} illustrates one such state transition. The reward is terminal-only:
\begin{equation}
\label{eq:reward}
\mathcal{W}(\mathbf{x}_v, \mathcal{P}_v,\mathbf{x}_{v+1})
=
\alpha \cdot \mathbb{I}(\text{all processes scheduled})
\cdot \left(\frac{OPT}{Obj}\right),
\end{equation}
where $Obj$ is the final objective value, $OPT$ is a time-capped lower bound returned by OR-Tools CP-SAT, and $\alpha>0$ is a scaling factor.

\textbf{Reward design and choice of DQN.}
The reward is terminal-only because the true objective depends on the final peak usage $R_k^{(V)}=\max_t u_k^{(V)}(t)$, so intermediate rewards would misalign with the optimization target. The ratio $OPT/Obj$ normalizes instance scale while remaining bounded in $(0,\alpha]$. Here $OPT$ is used only as an instance-level normalization constant. For a fixed training instance, the same time-capped lower bound is shared by all trajectories; hence it does not change the relative ordering of trajectories induced by their final objective values. A loose lower bound may rescale the rewards of that instance and thus affect the relative weighting of different instances in replay, but it does not make an inferior trajectory preferable within the same instance. In practice, this effect is mitigated by the bounded ratio reward, by tuning the scaling factor $\alpha$, and by evaluating the learned policy on held-out Normal and Hard instances. Moreover, the lower bound is required only during offline training and is not used during inference. We choose DQN because the action space is discrete and enumerable, the horizon is modest on the training split ($n\le 100$ on Easy), and terminal rewards pair naturally with replay-based value learning. \S~\ref{subsec:rl_algos} compares this choice against PPO and SAC.

\subsection{Candidate Local-Schedule Generation}
\label{sec:candidate_gen}

After the process-selection policy chooses \(\mathcal{P}_v\), iScheduler constructs the local subproblem \(\mathrm{RIP}_v\) exactly as defined in \S~\ref{sec:subproblem}. 
Candidate generation is performed by enumerating feasible schedules within the same local feasible region. 
Specifically, all candidates satisfy the same precedence constraints, time-window constraints, and residual-capacity constraints; we do not relax constraints or use different objective functions to create alternative candidates.

In both training and inference, we use the same CP-SAT configuration and collect up to \(m\) feasible schedules under a subproblem time cap
$\tau_{\mathrm{sub}} = 0.05 \times |T_{\mathcal{P}_v}|$.
Two candidates are treated as distinct if at least one task in \(T_{\mathcal{P}_v}\) receives a different start time. 
The solver first returns feasible solutions through capped solution collection. 
If fewer than \(m\) distinct candidates are found, we perturb the variable ordering and re-solve under the same constraints and time cap. 
If the pool is still smaller than \(m\), we pad it with the best feasible schedule found so far, which preserves feasibility and only creates duplicate inputs for the selector. 
We fix \(m=2\) in all reported experiments. 
Therefore, the learned selector operates on alternatives generated from the same constrained subproblem in both training and inference, and its advantage comes from choosing among feasible local commitments rather than from modifying the local optimization model.

\subsection{Learned Local-Solution Selector}
\label{subsec:solution-selection}

Given the candidate pool $\{A_{1,v},\dots,A_{m,v}\}$ produced by \S~\ref{sec:candidate_gen}, the selector picks which candidate to commit. We introduce a value network $F_{\phi}(x_v, A_{j,v})$ that estimates the final objective value when committing solution $A_{j,v}$ at this step; the selected solution is $A_{j^*,v}$ with $j^* = \arg\min_{j} F_{\phi}(x_v, A_{j,v})$.

\textbf{Feature design.}
We reuse the process-level graph and RPU as the global context. A GNN encoder produces a graph embedding $h_v$, and each candidate $A_{j,v}$ is represented by a local feature vector $z_{j,v}$ summarizing its incremental resource profile, overlap with neighboring processes, and temporal slack. The value network scores each candidate via
\begin{equation}
    s_{j,v} = F_{\phi}(x_v, A_{j,v})
    = \mathrm{MLP}\big([h_v \,\Vert\, z_{j,v}]\big),
\end{equation}
where $[\,\cdot\,\Vert\,\cdot\,]$ denotes vector concatenation.

\textbf{Offline training.}
We train $F_{\phi}$ offline from solved trajectories. For each state $x_v$ and candidate set $\{A_{j,v}\}_{j=1}^m$, we complete the remaining schedule with a fixed baseline policy (the current DQN policy) and record the resulting objective values $\{\mathrm{Obj}_{j,v}\}_{j=1}^m$. Rather than regressing absolute costs, we train with a pairwise ranking loss:
\begin{equation}\label{eq:rank_loss}
    \mathcal{L}_\mathrm{rank} =
    \sum_{\mathrm{Obj}_{j,v} < \mathrm{Obj}_{k,v}}
    \log\Big(1 + \exp\big(F_{\phi}(x_v, A_{j,v}) - F_{\phi}(x_v, A_{k,v})\big)\Big).
\end{equation}
This objective focuses on the relative quality of candidates and is robust to scale differences across instances.

\subsection{Training and Execution}
\label{subsec:rl-training}

Both learned components share the same reconfiguration-aware interface. Algorithm~\ref{alg:algorithm} summarises inference for the continual-optimization mapping $f(q,(S_i)_{i\in T},q')\!\rightarrow\!(S_i)_{i\in T'}$.

\begin{algorithm}[tb]
\caption{Continual Optimization Interface $f(q,(S_i)_{i\in T},q') \rightarrow (S_i)_{i\in T'}$ (iScheduler Execution, Reconfiguration-Aware)}
\label{alg:algorithm}
\textbf{Input}: Original RIP instance
$q = \langle T, \mathcal{R}, P, r, c, d, e, l\rangle$,
previous schedule $(S_i)_{i\in T}$,
updated instance $q' = \langle T', \mathcal{R}, P', r', c, d', e', l'\rangle$,
trained iScheduler agent with Q-function $Q$ and value network $F_{\phi}$\\
\textbf{Output}: Updated schedule $(S_i)_{i\in T'}$
\begin{algorithmic}[1]
\STATE $v \gets 0$ \COMMENT{Iteration counter}
\STATE $(G_{PL}, (u_{k})_{k\in \mathcal{R}}, \mathcal{C}_{PL})
       \gets \mathrm{InitializeReconfig}\big(q,(S_i)_{i\in T}, q'\big)$
       \COMMENT{Build $G_{PL}$ for $q'$ and reuse unchanged schedules}
\WHILE{$\mathcal{C}_{PL} \neq \emptyset$}
    \STATE $x_v \gets \mathrm{MDP}\big(G_{PL}, (u_{k})_{k\in \mathcal{R}}, \mathcal{C}_{PL}\big)$
    \STATE $\mathcal{P}_v \gets \arg\max_{\mathcal{P} \in \mathcal{C}_{PL}} Q(x_v,\mathcal{P})$
    \STATE $\mathrm{RIP}_v \gets \mathrm{Subproblem}(\mathcal{P}_v, (u_{k})_{k\in \mathcal{R}})$
    \STATE Solve $\mathrm{RIP}_v$ to obtain $m$ local solutions
           $\{A_{j,v}\}_{j=1}^m$
    \STATE Compute scores $s_{j,v} \gets F_{\phi}(x_v, A_{j,v})$ for $j=1,\dots,m$
    \STATE $j^* \gets \arg\min_{j} s_{j,v}$, commit
           $A_{j^*,v}$ to update $(u_{k})_{k\in \mathcal{R}}$
           and the schedule of tasks in $T_{\mathcal{P}_v}$
    \STATE Remove $\mathcal{P}_v$ from $\mathcal{C}_{PL}$
    \STATE Update $G_{PL}$ using $(u_{k})_{k\in \mathcal{R}}$
    \STATE $v \gets v + 1$
\ENDWHILE
\STATE Aggregate per-process schedules (including unchanged processes) to obtain $(S_i)_{i\in T'}$
\STATE \textbf{return} Updated schedule $(S_i)_{i\in T'}$
\end{algorithmic}
\end{algorithm}

At inference, iScheduler reuses unchanged process schedules, marks only affected processes as unscheduled, and then repeatedly selects a process by $\arg\max_{\mathcal{P}} Q(x_v,\mathcal{P})$, solves its local subproblem, scores the candidate local schedules with $F_\phi$, commits the best candidate, and updates the state until $\mathcal{C}_{PL}$ is empty. For training, we use a reconfiguration-aware DQN with experience replay and a periodically updated target network; at each step, the agent follows an $\epsilon$-greedy policy, stores the transition $(x_v,\mathcal{P}_v,r_v,x_{v+1})$, and optimizes the Bellman loss. The local-solution selector $F_\phi$ is trained separately and offline with the ranking loss in Eq.~\eqref{eq:rank_loss}. Exploration is used only during offline training; inference does not take exploratory actions. Training hyperparameters are deferred to \S~\ref{subsec:performance}.

\subsection{Discussion: Assumptions and Faithfulness}
\label{sec:discussion}

iScheduler combines decomposition, RL-guided process ordering, and iterative local commitment. 
We do not claim a worst-case approximation ratio or convergence guarantee for the end-to-end pipeline. 
Instead, this subsection provides three structural observations that clarify why the method is effective in the intended workload regime and where its failure modes arise.

\textbf{Feasibility preservation.}
At iteration \(v\), the local subproblem \(\mathrm{RIP}_v\) is solved under the residual-capacity constraint
\[
u^{(v)}_k(t) + d^{(v)}_k(t) \le R^{(v)}_k,\quad \forall k\in \mathcal{R},\forall t .
\]
After committing the selected local schedule, the usage profile is updated by
\[
u^{(v+1)}_k(t)=u^{(v)}_k(t)+d^{(v)}_k(t).
\]
Therefore, if every committed local schedule satisfies its local residual-capacity constraint, the aggregated schedule satisfies the global resource-capacity constraint by construction. 
Since weak-component decomposition does not remove precedence edges, precedence feasibility is also preserved inside each process. 
Thus, iterative commitment affects optimality but not feasibility, provided each local subproblem is solved feasibly.

\textbf{When is decomposition faithful?}
Decomposition is most faithful when the process-level interaction graph \(G_{{PL}}\) is sparse and feasible-window overlap leaves enough temporal slack to absorb local commitments. 
In the ideal case \(\rho(G_{{PL}})=0\), no two processes have overlapping feasible windows, and the processes are independent after decomposition. 
The decomposed procedure can then recover the monolithic optimum because no local commitment changes the feasible region of any remaining process. 
As \(\rho(G_{{PL}})\) increases, the gap between decomposed and monolithic optimization depends on how much early commitments inflate the peak resource usage relative to a jointly optimized schedule.

\textbf{When does learned ordering help?}
Learned ordering is useful when different process orders induce materially different RPU trajectories. 
This effect becomes stronger when the number of processes grows, when \(G_{{PL}}\) contains many contention edges, and when resource costs are heterogeneous. 
In these cases, a fixed priority rule observes only a limited local criterion, whereas the learned policy conditions on the current graph state and RPU. 
The goal of the policy is therefore not to replace the exact solver, but to choose an order that exposes easier residual subproblems and avoids avoidable peak increases.

\textbf{How do local-commitment errors propagate?}
Local-commitment errors are one-sided. 
A committed process adds to the current usage profile and cannot later be removed unless reconfiguration explicitly marks it as affected. 
Thus, a poor early commitment may shrink the feasible region of later subproblems and increase the terminal capacity
$R_k^{(V)}=\max_t u_k^{(V)}(t)$.
However, it does not invalidate already satisfied constraints. 
The learned local-solution selector is designed to reduce this error propagation by preferring feasible candidates whose incremental resource profiles leave more favorable residual capacity for later processes.

\section{Experimental Evaluation}
\label{sec:experiments}

This section presents a comprehensive experimental evaluation of iScheduler.  
We conduct extensive experiments on the proposed  
L-RIPLIB dataset  to answer the following key questions:

\begin{itemize}[leftmargin=*, labelsep=0.5em]
    \item[\textbullet] \textbf{How does iScheduler perform compared with existing optimization methods?} We compare its large-scale solving quality and runtime against MIP, CP, COpter, and POP (\S~\ref{subsec:performance}).

    \item[\textbullet] \textbf{How effective is the learned scheduling-order strategy?} We evaluate whether iScheduler's process-selection policy outperforms standard heuristic orders (\S~\ref{subsec:ordering}).

    \item[\textbullet] \textbf{How does the proposed solution-selection mechanism influence performance?} We compare the learning-based selection against alternative feasible-schedule selection heuristics (\S~\ref{subsec:solution-selection_}).
\end{itemize}

\begin{figure}[!t]
    \centering
{\includegraphics[width=0.9\linewidth]{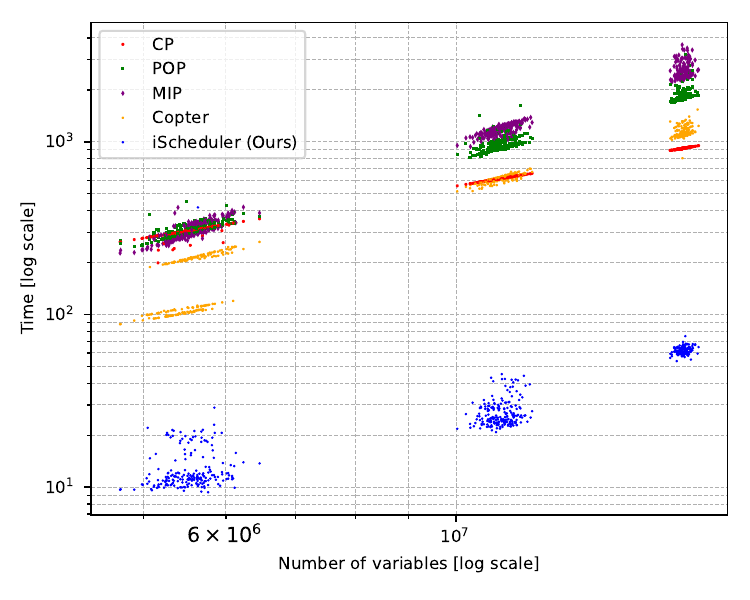}
}
\caption{Runtime versus problem size (number of variables, log scale). Each point corresponds to one test instance.}
\label{fig:compare}
\label{fig_result2}
\end{figure}

\begin{figure*}[!t]
    \centering
\subfloat[\textbf{Easy:} normalized cost vs. normalized runtime.]{\includegraphics[width=0.32\linewidth]{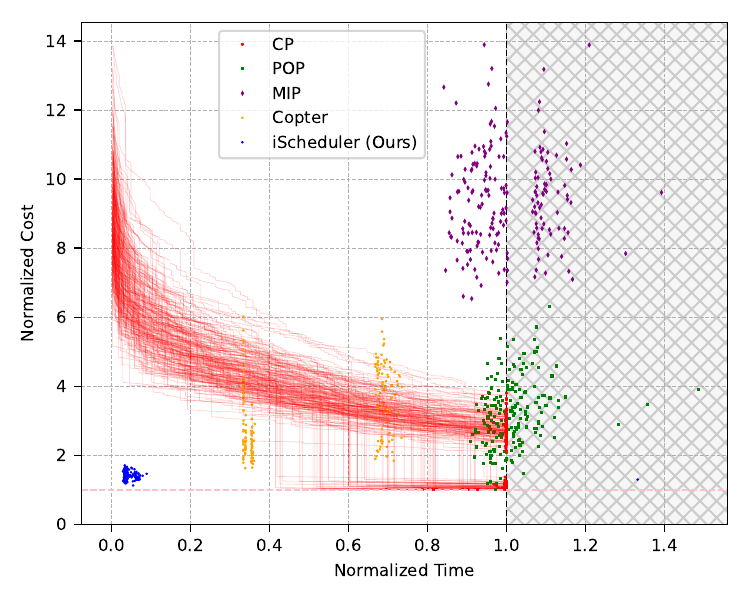}%
\label{fig_100}}
\hfill
\subfloat[\textbf{Normal:} normalized cost vs. normalized runtime.]{\includegraphics[width=0.32\linewidth]{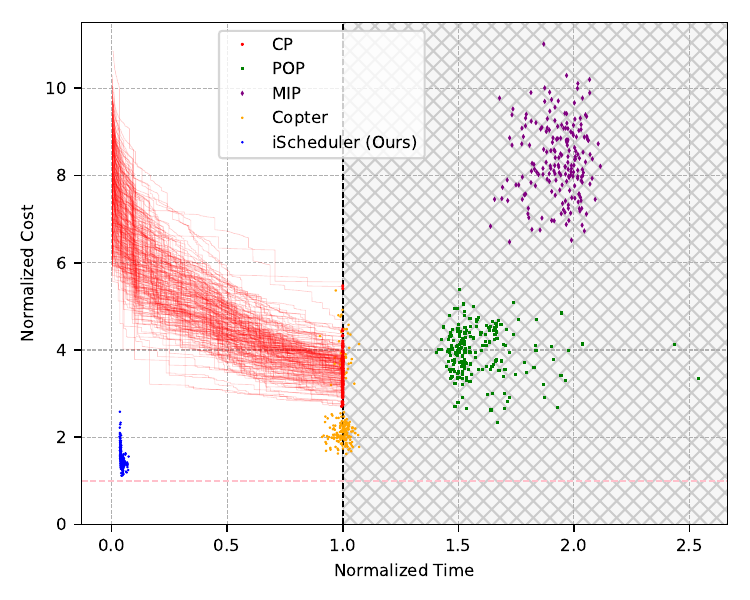}%
\label{fig_200}}\hfill
\subfloat[\textbf{Hard:} normalized cost vs. normalized runtime.]{\includegraphics[width=0.32\linewidth]{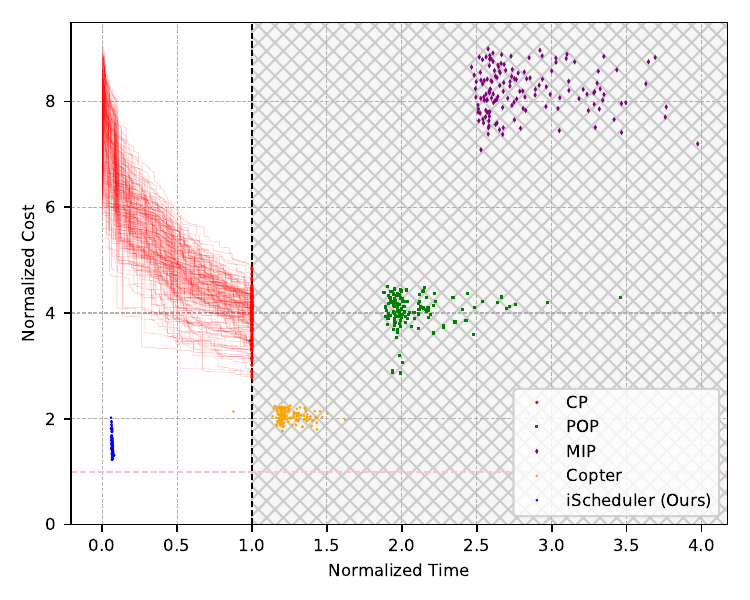}%
\label{fig_300}}
\caption{\textbf{Per-instance cost--time trade-offs on L-RIPLIB.}
Subfigures (a)--(c) visualize per-instance normalized behavior. For each instance, the y-axis reports the normalized objective $\textit{Obj}/\textit{LB}$, where $\textit{LB}$ is the best-known time-capped lower bound returned by OR-Tools CP-SAT, and the x-axis reports the normalized runtime $t/\mathrm{Time}_{\mathrm{limit}}$.
This normalization is used only to place instances with different objective scales and time budgets in the same plot.
Each marker corresponds to the final solution returned by iScheduler, MIP, COpter, and POP (best solution within the time budget, or the first feasible one if no feasible solution is found within $\mathrm{Time}_{\mathrm{limit}}$).
Only CP (OR-Tools CP-SAT) additionally plots its anytime improvement trajectory of incumbents over time.
}
\label{fig_result}
\end{figure*}

\subsection{Setup}

\begin{table}[t]
\footnotesize
\centering
\caption{Parameter values of the seed data.}
\label{table2}
\begin{tabular}{c|c}
\toprule
    Parameters & Values \\
    \midrule
    $n$ & $[100,1000]$\\
    $|T|$ &$[1000,40000]$ \\
    $|\mathcal{R}|$ &$\{1,3\}$\\
    $|P|$ &$[1931,36010]$\\\bottomrule
\end{tabular}
\end{table}
\begin{table}[t]
\footnotesize
\centering
\caption{Statistics of the L-RIPLIB benchmark.}
\label{table1}
\begin{tabular}{c|ccccc}
\toprule
  Category&\multicolumn{1}{c}{$n$
} & \multicolumn{1}{c}{ $|T|$}
   & \#Instances & Training&Testing\\
\midrule
 Easy&$100$& $[2634,3589]$&600 & 400&200\\
  Normal&$200$&  $[5566,6572]$&200 & 0&200\\
  Hard&$300$&  $[8929,9509]$&200 & 0&200\\\bottomrule
\end{tabular}
\end{table}

\begin{table*}[t]
\footnotesize
\centering
\caption{Performance comparison of iScheduler, MIP, CP, COpter, and POP on L-RIPLIB. The table reports raw, unnormalized ObjVal and wall-clock Time (in seconds), aggregated by mean and standard deviation over each split. Lower ObjVal and Time indicate better performance. If no feasible solution is found within the nominal time limit, Time denotes the wall-clock time to the first feasible solution and ObjVal denotes that first-feasible objective.}
\label{tab:performance}
\begin{tabular}{l|rr|rr|rr}
\toprule
\textbf{Method} &
\multicolumn{2}{c|}{\textbf{L-RIPLIB (Easy)}}&
\multicolumn{2}{c|}{\textbf{L-RIPLIB (Normal)}}&
\multicolumn{2}{c}{\textbf{L-RIPLIB (Hard)}}\\
 &
   \multicolumn{1}{c}{ObjVal}&\multicolumn{1}{c|}{Time}&
   \multicolumn{1}{c}{ObjVal} &\multicolumn{1}{c|}{Time}&
   \multicolumn{1}{c}{ObjVal} &\multicolumn{1}{c}{Time}\\
\midrule
MIP &11784.5$\pm$1810.6&309.65$\pm$41.47&20618.9$\pm$2218.7&1186.66$\pm$84.59&30240.1$\pm$1694.3&2597.03$\pm$306.21\\

CP &
2775.6$\pm$1089.4&304.63$\pm$20.73&
8809.3$\pm$1038.5& 613.18$\pm$18.25&
14401.2$\pm$1653.5&919.26$\pm$12.05\\

COpter &4078.2$\pm$1356.0 &157.38$\pm$57.96&6055.1$\pm$2015.0&612.87$\pm$30.87&7560.0$\pm$423.6&1150.01$\pm$86.65\\

POP-8&
4024.3$\pm$1200.0& 311.73$\pm$29.97&
9751.9$\pm$1395.6& 974.36$\pm$99.90&
14957.6$\pm$1123.1& 1907.02$\pm$212.14\\

iScheduler &
\textbf{1788.0$\pm$180.9}&\textbf{14.89$\pm$28.73}&
\textbf{3762.7$\pm$610.0}&\textbf{27.18$\pm$5.05}&
\textbf{5469.8$\pm$613.6}& \textbf{62.11$\pm$3.00}\\
\bottomrule
\end{tabular}
\end{table*}

\textbf{L-RIPLIB Dataset.} We build L-RIPLIB from 30 seed schedules collected from daily scheduling operations on a cloud computing platform. From these seeds, we generate 1,000 instances by varying four drivers: the number of processes $n$, the number of tasks $|T|$, the number of resources $|\mathcal{R}|$, and the number of precedence constraints $|P|$. Table~\ref{table2} summarizes the statistics of the seed data.

Each instance provides (i) the original problem definition $q$, (ii) an available feasible schedule $(S_i)_{i\in T}$ for $q$, and (iii) an updated instance $q'$ for reconfiguration. We obtain $q'$ by modifying 5\% of task parameters relative to $q$. We also report the CP-SAT solving time under a fixed nominal cap and the corresponding lower bound for each instance. As shown in Table~\ref{table1}, we stratify instances into \textit{Easy}, \textit{Normal}, and \textit{Hard} based on $n$, and we list the training/testing split used in our evaluation. Notably, we \emph{intentionally} train iScheduler only on \textit{Easy}; \textit{Normal/Hard} are held out for zero-shot OOD evaluation, which tests whether the learned policy transfers to larger-scale and harder regimes without access to Normal/Hard training data. Appendix~A gives the full schema and construction details of L-RIPLIB.

\textbf{Hardware and Environment.}
  All experiments run on a 2-socket AMD EPYC
7763 64-core processor with 512GB of DDR4 memory. We limit solvers (Gurobi 13.0.0~\cite{gurobi} and OR-Tools CP-SAT 9.14~\cite{cpsatlp}) to 8 threads (no significant speedup is observed beyond 8 threads~\cite{teal}).

\textbf{Baseline warm-starts.}
Because the evaluation is performed in the reconfiguration setting, each test case supplies the previous schedule $(S_i)_{i\in T}$ for the original instance $q$ alongside the updated instance $q'$. To make the comparison against iScheduler's partial rescheduling as fair as possible, every baseline is given access to $(S_i)_{i\in T}$ whenever its solver interface supports warm-starting, using the schedule entries of tasks that exist in both $q$ and $q'$ as the initial solution. Concretely: (i) \textbf{MIP} is warm-started by passing $(S_i)_{i\in T\cap T'}$ as a MIP start to Gurobi. (ii) \textbf{CP} is warm-started via OR-Tools CP-SAT's \texttt{AddHint} interface on the same task set. (iii) \textbf{COpter}~\cite{copter} uses warm-starts by construction: its Proximal Point Algorithm reuses the previous solution as the starting point. (iv) \textbf{POP}~\cite{narayanan2021solving} re-partitions the updated instance from scratch, consistent with the original POP formulation that does not define a reconfiguration-aware partition re-use; this mirrors the only fair way to evaluate the method as published. In our RIP adaptation of POP, the partition unit is a process: the process set is divided into 8 groups, and each group is solved as a RIP subproblem with the original constraints within the group preserved. After the partitioned solves, the resulting task start times are aggregated into a complete schedule. (v) \textbf{iScheduler} retains the schedules of unchanged processes exactly (Algorithm~\ref{alg:algorithm}) and reschedules only affected ones. All methods thus receive the strongest reconfiguration-aware initialization their respective formulations admit.

\textbf{Evaluation metrics and timeout protocol.}
All methods are evaluated under the same nominal time budget,
\(\mathrm{Timelimit}=0.1\times |T|\) seconds, and under the strongest warm-start condition supported by each method's formulation. 
We report two metrics: 
(i) \emph{ObjVal}, the total resource investment cost defined in Eq.~\eqref{eq1}, and 
(ii) \emph{Time}, the wall-clock time required to obtain the reported solution.

For each method and each instance, the reported objective and time always refer to the same feasible schedule. 
If a method finds at least one feasible schedule within $\mathrm{Time}_{\mathrm{limit}}$, we report its best feasible incumbent within the budget and the time at which that incumbent is obtained. 
If a method finds no feasible schedule within $\mathrm{Time}_{\mathrm{limit}}$, we continue the run until its first feasible schedule is found, and report that first-feasible objective together with the corresponding wall-clock time. 
Thus, no objective value is reported for an infeasible incumbent, and the ObjVal and Time columns in Table~\ref{tab:performance} are paired consistently.

This protocol is conservative for iScheduler's time-to-feasibility comparison: baselines that fail within the nominal budget are still allowed to continue until feasibility, rather than being assigned an infinite time or excluded from objective comparison. 
For reconfiguration, MIP and CP use the supplied previous schedule through their native warm-start interfaces, namely Gurobi MIP starts and OR-Tools CP-SAT hints. COpter uses its native warm-start mechanism, POP re-partitions the updated process set according to its original partitioned-optimization formulation, and iScheduler retains unchanged process schedules and reschedules only affected processes. Therefore, each method is evaluated under the strongest reconfiguration-aware initialization supported by its own formulation, and all final schedules are scored by the same RIP objective in Eq.~\eqref{eq1}.

\textbf{Training Details and Hyperparameter Search.}  
We train the reconfiguration-aware DQN agent following the procedure in \S~\ref{subsec:rl-training} on the training split of L-RIPLIB (Table \ref{table1}), using Adam ($lr=0.001$), replay buffer size $N=10,000$, minibatch size $k=128$, $\epsilon$ = 0.05, target update frequency $K$ = 10.
We tune $\alpha \in \{10, 100, 300, 500, 1000\}$   and $\gamma \in \{0, 0.8, 0.9, 0.95, 0.99\}$, selecting $\alpha=100$ , $\gamma=0.9$. For solution selection, we fix the pool size 
$m=2$ to limit enumeration overhead. 

\subsection{Performance Comparison}
\label{subsec:performance}

We evaluate all methods on L-RIPLIB in the \emph{reconfiguration} setting, where each test case provides an original instance $q$, an available schedule $(S_i)_{i\in T}$, and an updated instance $q'$ to be rescheduled under changed parameters.
We compare iScheduler against four strong baselines: (i) \textbf{MIP}: An MIP formulation solved by Gurobi; (ii) \textbf{CP}: A CP formulation solved using OR-Tools CP-SAT; (iii) \textbf{COpter}~\cite{copter}: A proximal point algorithm and warm-start-based continual optimization framework solved by Gurobi; (iv) \textbf{POP}~\cite{narayanan2021solving}: A partitioned-optimization baseline adapted to RIP by dividing the process set into 8 groups.

On large-scale instances (L-RIPLIB Normal/Hard), some baselines do not return a feasible schedule within the time limit, whereas iScheduler remains stable.
Across Easy/Normal/Hard, iScheduler reduces time-to-feasibility by $10.57\times$--$43.66\times$ (Table~\ref{tab:performance}) while preserving competitive solution quality.  Figure~\ref{fig:compare} summarizes the scalability trend: as the problem size (number of variables, on a log scale) increases, iScheduler consistently achieves the lowest runtime among all compared methods.
Figure~\ref{fig_result} further visualizes the cost-time frontier, where iScheduler consistently achieves a stronger balance between objective value and runtime, with the advantage being most pronounced on large and tightly constrained instances. The apparent difference between Figure~\ref{fig_result} and Table~\ref{tab:performance} comes from the visualization protocol: Figure~\ref{fig_result} normalizes each instance by its own CP-SAT lower bound and plots per-instance points/CP anytime trajectories, while Table~\ref{tab:performance} reports raw unnormalized ObjVal averaged over each split.

\subsection{Ablation Study}
\label{sec:ablation}

We randomly sample 20 instances from L-RIPLIB Hard and conduct ablations to isolate the impact of key design choices.
Unless otherwise stated, all ablated variants share the same decomposition pipeline and subproblem solver; differences arise only from the corresponding selection policy.

\begin{table}[t]
\footnotesize
\centering
\caption{Performance of different scheduling-order strategies on L-RIPLIB. Lower ObjVal and Time (in seconds) indicate better performance.}
\label{tab:scheduling_order}
\begin{tabular}{lcc}
\toprule
\textbf{Method} & \textbf{ObjVal} & \textbf{Time} \\
\midrule
CCPM & 5006.4$\pm$436.9& 57.96$\pm$4.73\\
MRRR & 5096.5$\pm$296.8 & 59.55$\pm$2.71\\
DUM  & 6055.4$\pm$276.2 & 57.86$\pm$4.12\\
RAND & 5965.7$\pm$411.6 & \textbf{57.25$\pm$3.25}\\
iScheduler & \textbf{4576.1$\pm$180.4} & 60.10$\pm$4.21\\
\bottomrule
\end{tabular}
\end{table}

\subsubsection{Comparison of Scheduling Orders}
\label{subsec:ordering}

We compare iScheduler's learned process-selection policy with classical ordering heuristics: (i) \textbf{Custom Critical Path Method (CCPM)}~\cite{10.1609/icaps.v33i1.27244}: Orders processes lexicographically by attributes $(LF_\mathcal{P}, PT_\mathcal{P})$, where $LF_\mathcal{P}$ is the latest finish time and $PT_\mathcal{P}$ is the total processing time; (ii) \textbf{Max Resource Requirement Rule (MRRR)}~\cite{10.1145/3449726.3463154}: A greedy heuristic that prioritizes scheduling the unscheduled process with the highest resource requirement in $\mathcal{C}_\mathrm{PL}$; (iii) \textbf{Dummy (DUM)}: A greedy heuristic that selects the unscheduled process with the largest task set size $|T_{\mathcal{P}}|$; (iv) \textbf{Random Ordering (RAND)}: Randomly selects a process from $\mathcal{C}_\mathrm{PL}$ at each iteration. 
All methods use the same decomposition and subproblem solver, so the only degree of freedom is the selection order.

As reported in Table~\ref{tab:scheduling_order}, iScheduler achieves the lowest objective value, reducing the average cost by 8.6-24.4\%, while incurring only minor overhead (within 5\% of the fastest heuristic RAND).
These results suggest that the learned policy captures long-range interaction patterns that fixed rules do not exploit.

\begin{table}[t]
\footnotesize
\centering
\caption{Objective value ratio (lower is better) of solution-selection strategies 
normalized to iScheduler's learning-based method.}
\begin{tabular}{lc}
\toprule
\textbf{Selection Strategy} & \textbf{ObjVal} \\
\midrule
MAD & 1.12$\times$\\
TV & 1.15$\times$\\
MCT & 1.28$\times$\\
RAND-LS & 1.63$\times$\\
iScheduler  & \textbf{1.00$\times$}\\
\bottomrule
\end{tabular}
\label{tab:selection}
\end{table}

\subsubsection{Comparison of Solution Selection Policies}
\label{subsec:solution-selection_}
We evaluate the learned ranking-based solution-selection module against representative heuristic policies: moving average of differences (MAD), total variation (TV), minimum completion time (MCT), and random local solution (RAND-LS).
Compared with these heuristic selection rules, iScheduler reduces the global cost by 10.7\%--38.7\% (Table~\ref{tab:selection}).
A plausible explanation is that smoothness-oriented metrics (e.g., MAD/TV) reduce local fluctuations but do not adequately account for cross-resource coupling, while MCT tends to commit aggressive early schedules that later restrict feasibility.
In contrast, the learned selector better aligns local commitments with downstream global impact.

\begin{table*}[t]
\footnotesize
\centering
\caption{Disentangling the three mechanisms of iScheduler on 20 L-RIPLIB Hard instances. Each row enables one additional mechanism on top of the previous row, isolating its marginal gain over decomposition alone. Lower ObjVal and Time (in seconds) indicate better performance; signed $\Delta$ columns report relative change over the decomposition-only baseline (a).}
\label{tab:disentangle}
\begin{tabular}{lcccc}
\toprule
\textbf{Variant} & \textbf{ObjVal} & \textbf{$\Delta$ObjVal} & \textbf{Time} & \textbf{$\Delta$Time} \\
\midrule
(a) Decomposition only (RAND+RAND-LS) & 14070.5$\pm$774.8 & --       & \textbf{15.91$\pm$0.38}   & -- \\
(b) iScheduler DQN for process order + RAND-LS & 8668.9$\pm$476.3 & $-38.4\%$ & 36.54$\pm$4.34 & $+1.3\times$ \\
(c) iScheduler & \textbf{4576.1$\pm$180.4} & \textbf{$-67.5\%$} & 60.10$\pm$4.21 & $+2.8\times$ \\
\bottomrule
\end{tabular}
\end{table*}

\subsubsection{Disentangling the Three Mechanisms}
\label{subsec:disentangle}

The preceding ablations isolate the two learned policies under controlled settings: 
Table~\ref{tab:scheduling_order} fixes the decomposition pipeline and local solver to test process ordering, while Table~\ref{tab:selection} fixes the process order and compares local-solution commitment rules. 
We further report a staged end-to-end ablation in Table~\ref{tab:disentangle} to clarify how the three mechanisms contribute to the complete system.

The three mechanisms affect different performance dimensions. 
Decomposition is primarily a tractability mechanism: it reduces each solver invocation from the monolithic RIP to a much smaller local subproblem, which explains the large reduction in time-to-feasibility reported in Table~\ref{tab:performance}. 
Given this decomposition, learned process ordering improves coordination across subproblems by choosing a more favorable RPU trajectory. 
Finally, learned local-solution selection controls the immediate resource-profile shape committed at each iteration, and therefore has a direct effect on the final peak resource capacity.

In the staged ablation, adding the learned process-ordering policy reduces the average resource cost over the decomposition-only variant, and replacing the unlearned local-solution rule with the learned selector yields the largest additional reduction. 
This result should not be read as a universal percentage attribution, because the marginal gain of each component depends on the strength of the other fixed components. 
Rather, it shows that iScheduler's performance is not explained by decomposition alone: the learned ordering and learned local-solution selector provide complementary improvements after the problem has been made tractable by decomposition.

\begin{table}[t]
\footnotesize
\centering
\caption{Comparison of RL algorithms for process selection on 20 L-RIPLIB Hard instances.
}
\label{tab:rl_algos}
\begin{tabular}{lccc}
\toprule
\textbf{Algorithm} & \textbf{Policy head} & \textbf{ObjVal} & \textbf{Time (s)} \\
\midrule
DQN (Ours)        & $Q(x,\mathcal{P})$, $\epsilon$-greedy & \textbf{4576.1$\pm$180.4} & \textbf{60.10$\pm$4.21} \\
PPO               & $\pi_\theta(\mathcal{P}\mid x)$ + $V$  & 4920.0$\pm$508.7          & 93.03$\pm$3.40          \\
SAC    & $\pi_\theta(\mathcal{P}\mid x)$ + $Q_{1,2}$ & 4871.3$\pm$297.8     & 65.64$\pm$2.47        \\
\bottomrule
\end{tabular}
\end{table}

\subsubsection{Comparison of RL Algorithms}
\label{subsec:rl_algos}
\S~\ref{subsec:rl-training} adopts a DQN for process selection. This choice is motivated by three properties of our setting: the action space at each iteration is a {discrete} subset $\mathcal{C}_{PL}\subseteq V_{PL}$, episode lengths are modest ($n\!\le\!100$ process selections on L-RIPLIB Easy), and the terminal-only reward in Eq.~\eqref{eq:reward} is well matched to value-based bootstrapping with experience replay. For completeness and to address the concern of whether a more modern RL algorithm would perform differently, we re-implement the process-selection head with PPO (on-policy actor-critic) and a discrete variant of SAC (off-policy actor-critic with entropy regularisation), keeping every other component of iScheduler identical: same GATv2 encoder, same state/action/reward, and same exact-CP local solver. Table~\ref{tab:rl_algos} reports ObjVal and Time on the same 20 L-RIPLIB Hard instances used for the ablations above.

Overall, these ablations support the design choice of combining solver-backed decomposition with two learned decision layers: the decomposition layer provides scalability, the DQN policy improves cross-process coordination, and the learned selector improves the quality of local commitments.

\subsection{Generalization to Public Benchmarks}
\label{sec:public_bench}

To examine whether iScheduler remains compatible with benchmark instances outside the L-RIPLIB generation pipeline, we further evaluate it on the PSPLIB benchmark family~\cite{kolisch1997psplib}, including the standard J30, J60, and J120 subsets. These subsets contain 30, 60, and 120 non-dummy activities per instance, respectively. Compared with L-RIPLIB, PSPLIB instances are much smaller: the largest J120 instance contains 122 tasks after adding dummy source and sink nodes, whereas L-RIPLIB Hard contains up to 10,000 tasks. Moreover, PSPLIB precedence graphs are connected end-to-end, so weak-component decomposition produces a single process for each instance. As a result, the scalability advantage of decomposition and the benefit of learned process ordering are largely inactive in this setting. Therefore, this experiment is not intended to establish a new state of the art on PSPLIB against highly tuned exact solvers. Instead, it serves as an out-of-distribution sanity check: iScheduler should remain feasible and competitive on public benchmark distributions that differ substantially from L-RIPLIB.

\textbf{Setup.}
We use the same iScheduler agent as in Section~\ref{subsec:performance}, trained only on the Easy split of L-RIPLIB, without any retraining or fine-tuning on PSPLIB. Since PSPLIB instances are static single-shot scheduling problems rather than reconfiguration streams, we run iScheduler in fresh-solve mode: the initial schedule is empty and all processes are marked unscheduled. Each PSPLIB instance is recast as an RIP by retaining its precedence graph, task durations, and per-resource demands. We set the task completion bounds according to the reported project horizon in the PSPLIB instance file and assign unit cost to each renewable resource. The per-instance time budget is
$\mathrm{Time}_{\mathrm{limit}}=\max(10\,\mathrm{s},\,0.1\times |T|)$,
which avoids artificial timeouts on very small instances. 
CP-SAT is run with a single search worker, and Gurobi MIP is run with
\texttt{Threads=1} and a \(5\%\) MIP gap tolerance. The iScheduler local solver uses the
same CP-SAT configuration as in the main experiments.

\textbf{Results.}
Table~\ref{tab:psplib} reports the results. Across J30, J60, and J120, the three methods obtain similar objective values, with differences within approximately \(2\%\). CP-SAT remains highly competitive on J30 and J60, and iScheduler obtains slightly better objective values on J120. These results indicate that the solver-backed iScheduler pipeline remains robust on public benchmark instances that differ from its training distribution.

The runtime results show the opposite pattern. On PSPLIB, monolithic CP-SAT is substantially faster than iScheduler. This is expected because PSPLIB instances are small and connected, so weak-component decomposition degenerates to a single process and provides no meaningful reduction in subproblem size. In this case, iScheduler still incurs framework overhead from instance construction, graph encoding, and neural inference, while receiving little benefit from decomposition or learned process ordering. This behavior is consistent with the intended scope of the method: iScheduler is designed for large, multi-process industrial RIP instances, where the overhead is amortized across many subproblems and where learned ordering affects the evolution of resource-pool usage. On L-RIPLIB Hard, this regime is reached, and the time ranking is reversed, as shown in Table~\ref{tab:performance}.
\begin{table*}[t]
\footnotesize
\centering
\caption{Generalization to PSPLIB. iScheduler is trained only on L-RIPLIB Easy and evaluated zero-shot on the full standard subsets (J30: 480 instances, J60: 480, J120: 600). Lower ObjVal and Time (s) are better. All methods receive the same time budget $\mathrm{Time}_{\mathrm{limit}}=\max(10\,\mathrm{s},\,0.1\times|T|)$. The bold cell per subset marks the best ObjVal.}
\label{tab:psplib}
\begin{tabular}{l|rr|rr|rr}
\toprule
\textbf{Method} & \multicolumn{2}{c|}{\textbf{J30}} & \multicolumn{2}{c|}{\textbf{J60}} & \multicolumn{2}{c}{\textbf{J120}}\\
 & ObjVal & Time & ObjVal & Time & ObjVal & Time \\
\midrule
MIP         & 75.95$\pm$22.70 & 1.82$\pm$3.19 & 96.52$\pm$33.55 & 6.59$\pm$4.38   & 136.93$\pm$53.28 & 11.42$\pm$2.63 \\
CP          & \textbf{75.25$\pm$22.54} & \textbf{0.05$\pm$0.21} & \textbf{94.69$\pm$33.10} & \textbf{3.28$\pm$4.35}   & 135.14$\pm$57.73 & \textbf{8.94$\pm$5.17} \\
iScheduler  & 75.30$\pm$22.57 & 0.23$\pm$0.75 & 95.64$\pm$33.50 & 3.74$\pm$4.65 & \textbf{134.05$\pm$52.73} & 9.98$\pm$6.99 \\
\bottomrule
\end{tabular}
\end{table*}

\section{Limitations}
\label{sec:limitation}

Although iScheduler achieves strong empirical performance, several limitations remain.

\textbf{Dependence on process-level separability.} 
The current weak-component decomposition is designed to preserve precedence feasibility without introducing cross-process boundary constraints. This choice is well matched to large industrial RIP instances composed of multiple job chains or service groups, but it can degenerate to a single process when the task DAG is weakly connected. In such cases, the learned process-ordering policy becomes inactive, and the framework mainly behaves as a solver-backed scheduling wrapper with additional overhead. A natural future direction is to add a boundary-aware graph-partitioning layer for connected DAGs, where cross-partition precedence relations are represented explicitly and preserved during local optimization and iterative commitment.

\textbf{Training cost of the solution-selection module.}
The current solution-selection module evaluates multiple candidate local schedules at each iteration and trains a value network with a pairwise ranking objective. Constructing supervision requires completing the remaining scheduling procedure for each candidate to obtain its final cost, which increases the number of solver rollouts. Future work can reduce this cost through more selective candidate generation, lighter selector architectures, and off-policy reuse of solved trajectories.

\textbf{Limited formal performance characterization.}
iScheduler combines decomposition, reinforcement learning, and iterative local commitment. Section~\ref{sec:discussion} discusses why these components are expected to work well in the intended workload regime, but the current paper does not provide a worst-case approximation ratio or an end-to-end convergence guarantee. A stronger theoretical treatment would characterize how the optimality gap depends on the density of the process-level interaction graph, the amount of temporal overlap, and the fraction of tasks modified during reconfiguration.

\textbf{Serial iterative execution.}
The current implementation schedules one process at a time. This design matches the sequential MDP formulation and keeps comparisons against serial baselines fair, but it does not exploit potential parallelism among weakly interacting processes. A natural extension is conflict-aware parallel commitment, where non-adjacent or low-contention processes in \(G_{\mathrm{PL}}\) are solved and committed together, followed by a state update. Such a design could improve throughput on large instances with many independent or weakly coupled process groups.

\textbf{Lack of semantic high-level reasoning.}
The current framework relies on graph structure, resource profiles, and solver feedback, but it does not use semantic information such as task names, service ownership, operator intent, or domain-specific scheduling rules. Large language models (LLMs) could be useful above the solver and RL layers, for example, to suggest process groupings from metadata, initialize scheduling heuristics, or translate natural-language reconfiguration requests into parameter edits \cite{NEURIPS2025_31e89283}. We leave this direction to future work because the current L-RIPLIB instances do not contain the semantic metadata needed to evaluate such integration rigorously.

\textbf{Policy expressiveness and reward sparsity.} Although the state representation
combines the process-level graph with the global RPU profile, the learned policy
still relies on finite-depth message passing and summary features. It may miss
high-order global dependencies that only emerge across many weakly coupled
processes. In addition, the terminal-only reward aligns with the final resource
investment objective but creates a sparse credit-assignment signal. We mitigate
this issue through replay-based value learning and relatively short training
episodes, while richer reward shaping and global-context encoders remain
directions for future work.

\section{Conclusion}
This paper shows that large-scale RIPs benefit from being solved as a sequence of coupled subproblems rather than as a single monolithic program. We propose iScheduler, which models iterative decomposition as an MDP and learns process-selection policies to navigate long-range interactions induced by shared resources and overlapping time windows. To support realistic evaluation, we introduce L-RIPLIB, an industrial-scale benchmark with 2,500–10,000 tasks per instance.
Experiments demonstrate that iScheduler achieves competitive resource costs with substantially lower time-to-feasibility than strong solver baselines at scale. Under dynamic updates, iScheduler reuses unchanged schedules and reschedules only affected processes, which reduces reconfiguration latency while preserving feasibility and solution quality. Overall, these results support a learning-guided continual optimization approach for industrial-sized RIP instances.

\fontsize{8.5pt}{10pt}\selectfont

\section*{Acknowledgements}

The research is partially supported by Quantum Science and Technology-National Science and Technology Major Project (QNMP) 2021ZD0302900, China National Natural Science Foundation with No. 92567301, 62132018, 62231015, ``Pioneer'' and ``Leading Goose'' R\&D Program of Zhejiang, 2023C01029 and 2023C01143, and Anhui Provincial Science and Technology Innovation Program 202523o09050015 and 202523o09050019.

\section*{Competing interests}
The authors declare that they have no competing interests or financial conflicts to disclose.

\section*{Appendixes}

\subsection*{Appendix A L-RIPLIB Dataset}\label{Dataset}

The L-RIPLIB dataset is publicly available~\cite{lriplib-dataset}.
In L-RIPLIB, each instance is stored in a JSON format, which contains the following key elements:
\begin{itemize} [leftmargin=*, labelsep=0.5em]
\item Tasks $T$: The list of task names, representing the various activities to be managed within the instance. 
\item Earliest\_start $e$: The list of earliest start times for each task, indicating the minimum time at which each task can commence. 
\item Deadline $l$: The list of deadlines (latest finish times) for each task.
\item Duration $d$: The list of durations for each task, specifying how long each task is expected to take from start to finish. 
\item Dependencies $P$: The list of task dependencies, outlining which tasks must be completed before others can begin, thereby establishing the sequence of execution. 
\item Resources $r$: The resources allocated to each task, detailing the specific inputs or tools required for task completion. 
\item Costs $c$: The unit cost of each kind of resource;
\item Task\_start $(S_i)_{i\in T}$: A solution given by CP-SAT within a limited time ($0.1\times|T|$ seconds).
\item Best\_cost: The total resource cost corresponding to the solution;
\item Time: The time CP-SAT used to solve this instance;
\item Bound: A lower bound of total resource cost given by CP-SAT;
\item Modified\_data $\Delta q$: the difference between $q$ and $q'$.
\end{itemize}

\begin{table*}[t]
\centering
\footnotesize
\caption{Symbols and Definitions.}\label{symbols}
\begin{tabular}{l l}
\toprule
\textbf{Symbol} & \textbf{Definition} \\
\midrule
$T$ & Set of tasks in a Resource Investment Problem (RIP) instance. \\
$\mathcal{R}$ & Set of renewable resource types. \\
$P$ & Set of precedence constraints, $P \subseteq T \times T$. \\
$t$ & Discrete time index in the scheduling horizon.\\
$n$ & Number of processes obtained after decomposition. \\
$q$ & An RIP instance (task set, resources, precedence, demands, costs, durations, earliest starts, deadlines). \\
$d_i$ & Duration (processing time) of task $i$. \\
$r_{i,k}$ & Demand of task $i$ for resource $k$ during its execution. \\
$c_k$ & Cost of provisioning one unit of resource $k$. \\
$R_k$ & Provisioned capacity (amount) of resource $k$ for the whole project. \\
$e_i$ & Earliest start time of task $i$. \\
$l_i$ & Deadline (latest completion bound) of task $i$ (used via $S_i \le l_i-d_i$). \\
$S_i$ & Scheduled start time of task $i$. \\
$\mathcal{U}(S,t)$ & Set of tasks active at time $t$ under schedule $S$. \\
$G_T=(T,P)$ & Task-level directed acyclic graph (DAG): nodes are tasks; edges are precedence constraints. \\
$\mathcal{P}_i$& A process (task subset) obtained by grouping tasks via weakly connected components of $G_T$. \\
$T_{\mathcal{P}_i}$& Set of tasks belonging to process $\mathcal{P}_i$.\\
$G_{PL}=(V_{PL},E_{PL})$ & Process-level interaction graph; nodes are processes; edges represent process interactions (e.g., overlapping feasible windows). \\
$v$ & Iteration index of the iterative scheduling procedure. \\
$\mathcal{P}_v$ & Process selected by the agent at iteration $v$ to schedule next. \\
$\mathrm{RPU}$ & Resource Pool Usage profile induced by already scheduled processes. \\
$u_k(t)$ & Current usage of resource $k$ at time $t$ from already committed (scheduled) processes. \\
$\mathrm{RIP}_v$ & Subproblem constructed at iteration $v$ for the selected process $\mathcal{P}_v$ given the current committed schedule/usage. \\
$R^{(v)}_k$ & Global provisioned capacity of resource $k$ after iteration $v$ (used in subproblem constraints together with $u_k(t)$). \\
$\mathcal{C}_{PL}$ & Candidate set of unscheduled processes in $G_{PL}$. \\
$m$ & Solution pool size (number of candidate local schedules generated per subproblem for solution selection). \\
$\alpha$ & Positive scaling factor used in the terminal reward. \\
$ES_P$ & Earliest start of a process: $ES_P=\min_{i\in T_P} e_i$. \\
$LF_P$ & Latest finish bound of a process: $LF_P=\max_{i\in T_P} l_i$. \\
$t_{\text{start}},t_{\text{end}}$ & Overlap window bounds for an edge $(P_i,P_j)$: $t_{\text{start}}=\max(ES_{P_i},ES_{P_j})$, $t_{\text{end}}=\min(LF_{P_i},LF_{P_j})$. \\
\bottomrule
\end{tabular}
\end{table*}
\subsection*{Appendix B Industrial Deployment and Use Cases of iScheduler}

The exponential growth of enterprise data and the shift toward data-driven decision-making have made efficient resource scheduling a business-critical capability.  Modern data-processing platforms—whether on-premises clusters or third-party clouds—submit thousands of heterogeneous jobs each day, driving highly variable demand for CPU, memory, and I/O resources.  Poorly timed job launches create pronounced load spikes, which in turn throttle throughput, delay downstream analytics, and inflate infrastructure costs.
\begin{figure}[t]
    \centering
    \includegraphics[width=0.47\textwidth]{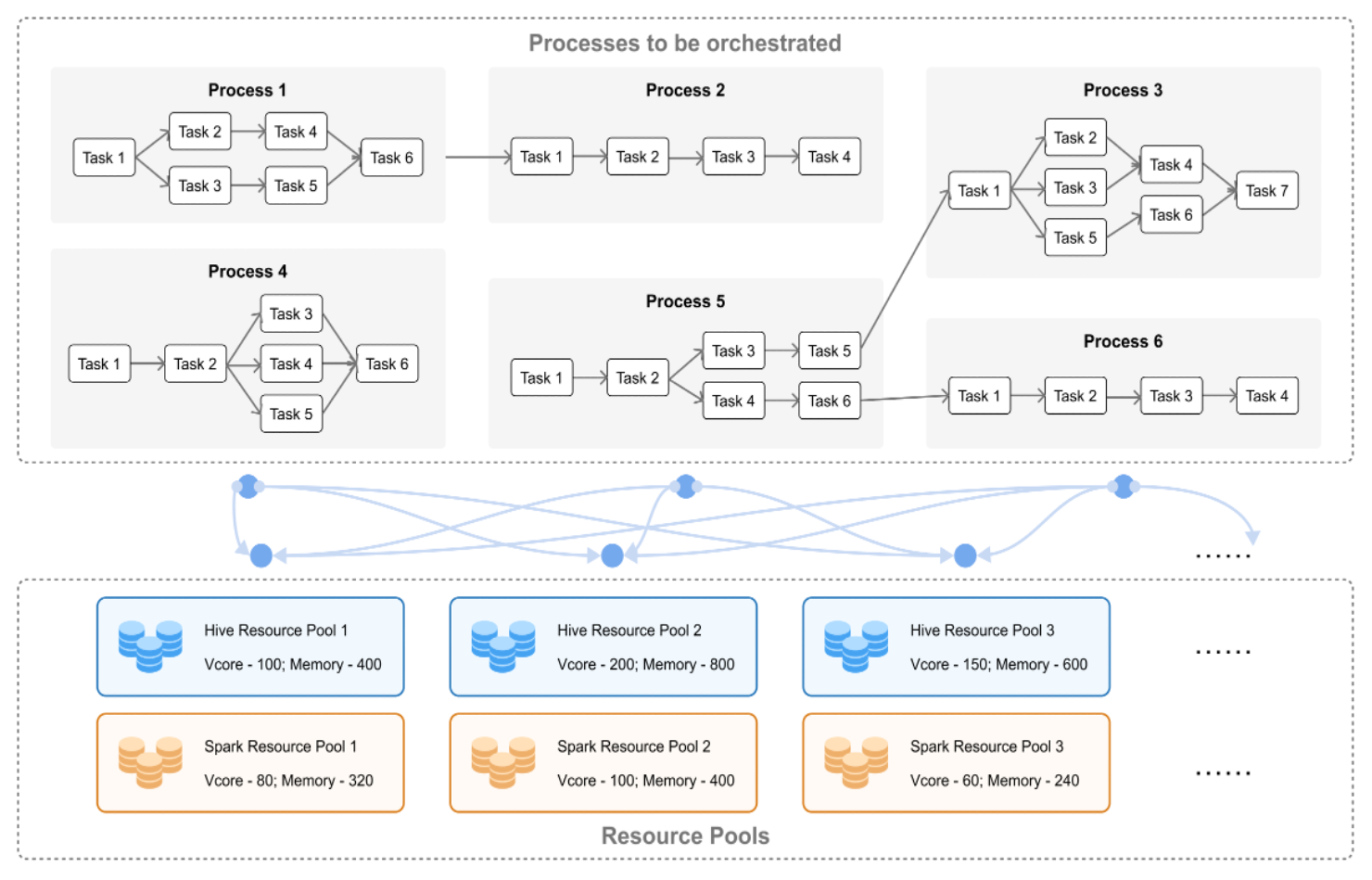}
    \caption{Synapse: Processes to be orchestrated and resource pools.}
    \label{fig:synapse}
\end{figure}
\textbf{Synapse production deployment.}
Figure~\ref{fig:synapse} depicts \textit{Synapse}, a fixed-quota data platform that previously relied on heuristic, expert-driven job calendars.  Before iScheduler, roughly half of the 24-hour horizon regularly entered a high-load regime, 
 during which jobs queued, timed out, or even failed, triggering costly re-runs.  By replacing manual calendars with iScheduler’s RL-guided iterative planning, Synapse now \emph{flattens} its CPU-utilization curve, shifting workloads from peak to off-peak windows, achieving {\$3.6\ M annual savings} in compute spend and {120 person-days} of orchestration effort eliminated per year.

From a data-center perspective, iScheduler acts as a temporal load-shaping mechanism that performs peak shaving and valley filling: it shifts deferrable workloads away from short-term peaks and improves infrastructure utilization. This matters because a substantial fraction of data-center electricity is associated with power delivery and cooling; empirical evidence reports that cooling alone accounts for about 40\% of total data-center energy consumption in certain deployments \cite{SONG20151255}. In parallel, the rapid scaling of LLM~\cite{GPT4,Llama3.1} training and inference introduces increasingly energy-intensive GPU workloads, and recent studies quantify the end-to-end carbon footprint of LLMs and show that inference/efficiency choices materially change total energy use \cite{faiz2024llmcarbon,fernandez-etal-2025-energy,strubell-etal-2019-energy}. Against this backdrop, reducing instantaneous peak power while preserving the required computation lowers the headroom required by power-and-cooling infrastructure and mitigates demand peaks, aligning iScheduler with cost- and sustainability-driven objectives in modern computing centers.

\textbf{Generalization beyond data pipelines.}
Because iScheduler optimizes RIP instances rather than a single application
domain, it can be adapted to other large-scale scheduling contexts.
\begin{itemize}[leftmargin=*, labelsep=0.5em]
   \item \textbf{Virtual Machine (VM) Rescheduling.}  
Cloud-scale data centers constantly create and retire VMs, leaving ``resource crumbs''—small, unusable CPU and memory slices—scattered across physical hosts. Periodic consolidation (i.e., migrating selected VMs to alternate hosts~\cite{10.1145/3689031.3717476}) allows idle machines to be powered down, but choosing which VMs to move and in what order is an RIP in its own right. By modelling each live-migration as a task with bandwidth and memory-copy constraints, iScheduler can generate a migration sequence that {minimises residual fragmentation}.

    \item \textbf{Development-Team Task Scheduling.}
Software-company dev teams often work on multiple project versions in parallel, each with distinct workloads and deadlines, while interdependent tasks from different functional units—UI, testing, and core development—must be coordinated~\cite{9223172,7965442}. Viewed as a complex RIP, this scenario can also be tackled by iScheduler: its iterative solving strategy produces a practical shift plan that guides the team’s day-to-day scheduling.
    
\end{itemize}

\subsection*{Appendix C Symbols and Definitions}
The symbols we used in this paper and corresponding definitions are listed in Table \ref{symbols}.

\bibliographystyle{fcs}
\bibliography{ref}

@article{1,
  title   = {Generating test data for both paths coverage and faults detection using genetic algorithms: multi-path case},
  author  = {Zhang, Yan and Gong, Dunwei},
  journal = {Frontiers of Computer Science in China},
  volume  = {8},
  number  = {5},
  pages   = {726--740},
  year    = {2014}
}

@article{2,
  title   = {Wolfhard {H G}. Flames. 2nd ed},
  author  = {Gaydon, A G},
  journal = {London: Chapman and Hall Ltd},
  year    = {1960}
}

@inproceedings{
faiz2024llmcarbon,
title={{LLMC}arbon: Modeling the End-to-End Carbon Footprint of Large Language Models},
author={Ahmad Faiz and Sotaro Kaneda and Ruhan Wang and Rita Chukwunyere Osi and Prateek Sharma and Fan Chen and Lei Jiang},
booktitle={The Twelfth International Conference on Learning Representations},
year={2024},
url={https://openreview.net/forum?id=aIok3ZD9to}
}

@inproceedings{fernandez-etal-2025-energy,
    title = "Energy Considerations of Large Language Model Inference and Efficiency Optimizations",
    author = "Fernandez, Jared  and
      Na, Clara  and
      Tiwari, Vashisth  and
      Bisk, Yonatan  and
      Luccioni, Sasha  and
      Strubell, Emma",
    editor = "Che, Wanxiang  and
      Nabende, Joyce  and
      Shutova, Ekaterina  and
      Pilehvar, Mohammad Taher",
    booktitle = "Proceedings of the 63rd Annual Meeting of the Association for Computational Linguistics (Volume 1: Long Papers)",
    month = jul,
    year = "2025",
    address = "Vienna, Austria",
    publisher = "Association for Computational Linguistics",
    url = "https://aclanthology.org/2025.acl-long.1563/",
    doi = "10.18653/v1/2025.acl-long.1563",
    pages = "32556--32569",
    ISBN = "979-8-89176-251-0"
}

@inproceedings{strubell-etal-2019-energy,
    title = "Energy and Policy Considerations for Deep Learning in {NLP}",
    author = "Strubell, Emma  and
      Ganesh, Ananya  and
      McCallum, Andrew",
    editor = "Korhonen, Anna  and
      Traum, David  and
      M{\`a}rquez, Llu{\'i}s",
    booktitle = "Proceedings of the 57th Annual Meeting of the Association for Computational Linguistics",
    month = jul,
    year = "2019",
    address = "Florence, Italy",
    publisher = "Association for Computational Linguistics",
    url = "https://aclanthology.org/P19-1355/",
    doi = "10.18653/v1/P19-1355",
    pages = "3645--3650"
}

@article{SONG20151255,
title = {Data Center Energy and Cost Saving Evaluation},
journal = {Energy Procedia},
volume = {75},
pages = {1255-1260},
year = {2015},
note = {Clean, Efficient and Affordable Energy for a Sustainable Future: The 7th International Conference on Applied Energy (ICAE2015)},
issn = {1876-6102},
doi = {https://doi.org/10.1016/j.egypro.2015.07.178},
url = {https://www.sciencedirect.com/science/article/pii/S1876610215009467},
author = {Z. Song and X. Zhang and C. Eriksson}
}

@misc{brody2021attentive,
  title={How attentive are graph attention networks?},
  author={Brody, Shaked and Alon, Uri and Yahav, Eran},
  howpublished={arXiv preprint arXiv:2105.14491},
  year={2021}
}

@misc{gurobi,
  author = {{Gurobi Optimization, LLC}},
  title = {{Gurobi Optimizer Reference Manual}},
  year = 2025,
  url = "https://www.gurobi.com"
}

@article{10.1609/icaps.v33i1.27244,
author = {Teichteil-K\"{o}nigsbuch, Florent and Pov\'{e}da, Guillaume and Barba, Guillermo Gonz\'{e}lez de Garibay and Luchterhand, Tim and Thi\'{e}baux, Sylvie},
title = {Fast and robust resource-constrained scheduling with graph neural networks},
year = {2023},
journal = {Proceedings of the Thirty-Third International Conference on Automated Planning and Scheduling (ICAPS)},
volume={32},
pages = {623--633},
}

@inproceedings{narayanan2021solving,
  title={Solving large-scale granular resource allocation problems efficiently with {POP}},
  author={Narayanan, Deepak and Kazhamiaka, Fiodar and Abuzaid, Firas and Kraft, Peter and Agrawal, Akshay and Kandula, Srikanth and Boyd, Stephen and Zaharia, Matei},
  booktitle={Proceedings of the ACM SIGOPS 28th Symposium on Operating Systems Principles},
  pages={521--537},
  year={2021}
}

@article{LI1992370,
title = {An iterative scheduling technique for resource-constrained project scheduling},
journal = {European Journal of Operational Research},
volume = {56},
pages = {370-379},
year = {1992},
author = {K.Y. Li and R.J. Willis},
}

@INPROCEEDINGS{9223172,
  author={Fireteanu, Vlad-Valentin},
  booktitle={2020 12th International Conference on Electronics, Computers and Artificial Intelligence (ECAI)}, 
  title={Agile Methodology Advantages when delivering Internet of Things projects}, 
  year={2020},
  volume={},
  number={},
  pages={1-5},
  doi={10.1109/ECAI50035.2020.9223172}}

@INPROCEEDINGS{7965442,
  author={Kroll, Josiane and Friboim, Shai and Hemmati, Hadi},
  booktitle={2017 IEEE/ACM 39th International Conference on Software Engineering: Software Engineering in Practice Track (ICSE-SEIP)}, 
  title={An Empirical Study of Search-Based Task Scheduling in Global Software Development}, 
  year={2017},
  volume={},
  number={},
  pages={183-192},
  doi={10.1109/ICSE-SEIP.2017.30}}

@inproceedings{10.1145/3689031.3717476,
author = {Ding, Xianzhong and Zhang, Yunkai and Chen, Binbin and Ying, Donghao and Zhang, Tieying and Chen, Jianjun and Zhang, Lei and Cerpa, Alberto and Du, Wan},
title = {Towards {VM} Rescheduling Optimization Through Deep Reinforcement Learning},
year = {2025},
isbn = {9798400711961},
publisher = {Association for Computing Machinery},
address = {New York, NY, USA},
url = {https://doi.org/10.1145/3689031.3717476},
doi = {10.1145/3689031.3717476},
booktitle = {Proceedings of the Twentieth European Conference on Computer Systems},
pages = {686–701},
numpages = {16},
location = {Rotterdam, Netherlands},
series = {EuroSys '25}
}

@article{kolisch1997psplib,
  title={{PSPLIB}-a project scheduling problem library: {OR} software-{ORSEP} operations research software exchange program},
  author={Kolisch, Rainer and Sprecher, Arno},
  journal={European Journal of Operational Research},
  volume={96},
  number={1},
  pages={205--216},
  year={1997},
  publisher={Elsevier}
}

@article{CAI2024102628,
title = {Deep reinforcement learning for solving resource constrained project scheduling problems with resource disruptions},
journal = {Robotics and Computer-Integrated Manufacturing},
volume = {85},
pages = {102628},
year = {2024},
author = {Hongxia Cai and Yunqi Bian and Lilan Liu},
}

@article{LIU2021107553,
title = {A three-stage decomposition algorithm for decentralized multi-project scheduling under uncertainty},
journal = {Computers and Industrial Engineering},
volume = {160},
pages = {107553},
year = {2021},
author = {Dongning Liu and Zhe Xu and Feifei Li},
}

@inproceedings{10.1145/3449726.3463154,
author = {Regnier-Coudert, Olivier and Pov\'{e}da, Guillaume},
title = {An empirical evaluation of permutation-based policies for stochastic {RCPSP}},
year = {2021},
url = {https://doi.org/10.1145/3449726.3463154},
doi = {10.1145/3449726.3463154},
booktitle = {Proceedings of the Genetic and Evolutionary Computation Conference Companion (GECCO)},
pages = {1451–1458},
location = {Lille, France}
}

@inproceedings{10.5555/3491440.3491681,
author = {Ganian, Robert and Hamm, Thekla and Mescoff, Guillaume},
title = {The complexity landscape of resource-constrained scheduling},
year = {2021},
booktitle = {Proceedings of the International Joint Conference on Artificial Intelligence (IJCAI)},
articleno = {241},
numpages = {7},
location = {Yokohama, Yokohama, Japan},
pages={1548-1554}
}

@article{doi:10.1287/opre.1060.0358,
  title={A decomposition-based genetic algorithm for the resource-constrained project-scheduling problem},
  author={Debels, Dieter and Vanhoucke, Mario},
  journal={Operations Research},
  volume={55},
  number={3},
  pages={457--469},
  year={2007},
  publisher={INFORMS}
}

@misc{cpsatlp,
  title = {{CP-SAT} v9.13},
  author = {Laurent Perron and Frédéric Didier},
  organization = {Google},
  year = {2025}
}

@article{KRETER2018472,
title = {Mixed-integer linear programming and constraint programming formulations for solving resource availability cost problems},
journal = {European Journal of Operational Research},
volume = {266},
number = {2},
pages = {472-486},
year = {2018},
issn = {0377-2217},
author = {Kreter, Stefan and Schutt, Andreas and Stuckey, Peter J. and Zimmermann, J{\"u}rgen},
}

@inproceedings{copter,
author = {Subramanya, Suhas Jayaram and Dennis, Don Kurian and Smith, Virginia and Ganger, Gregory R.},
title = {{COpter}: Efficient Large-Scale Resource-Allocation via Continual Optimization},
year = {2025},
isbn = {9798400718700},
publisher = {Association for Computing Machinery},
address = {New York, NY, USA},
url = {https://doi.org/10.1145/3731569.3764846},
doi = {10.1145/3731569.3764846},
booktitle = {Proceedings of the ACM SIGOPS 31st Symposium on Operating Systems Principles},
pages = {322–340},
numpages = {19},
location = {Lotte Hotel World, Seoul, Republic of Korea},
series = {SOSP '25}
}

@inproceedings{teal,
author = {Xu, Zhiying and Yan, Francis Y. and Singh, Rachee and Chiu, Justin T. and Rush, Alexander M. and Yu, Minlan},
title = {Teal: Learning-Accelerated Optimization of {WAN} Traffic Engineering},
year = {2023},
doi = {10.1145/3603269.3604857},
booktitle = {Proceedings of the ACM SIGCOMM 2023 Conference},
pages = {378–393},
numpages = {16},
}

@article{li2022scheduling,
  title={Scheduling multi-tenant cloud workflow tasks with resource reliability},
  author={Li, Xiaoping and Pan, Dongyuan and Wang, Yadi and Ruiz, Ruben},
  journal={Science China Information Sciences},
  volume={65},
  number={9},
  pages={192106},
  year={2022},
  publisher={Springer}
}

@article{liu2024muxflow,
  title={{MuxFlow}: efficient {GPU} sharing in production-level clusters with more than 10000 {GPU}s},
  author={Liu, Xuanzhe and Zhao, Yihao and Liu, Shufan and Li, Xiang and Zhu, Yibo and Liu, Xin and Jin, Xin},
  journal={Science China Information Sciences},
  volume={67},
  number={12},
  pages={222101},
  year={2024},
  publisher={Springer}
}

@article{GPT4,
  title={{GPT-4} Technical Report},
  author={Achiam, Josh and Adler, Steven and Agarwal, Sandhini and Ahmad, Lama and Akkaya, Ilge and Aleman, Florencia Leoni and Almeida, Diogo and Altenschmidt, Janko and Altman, Sam and Anadkat, Shyamal and others},
  journal={arXiv preprint arXiv:2303.08774},
  year={2023}
}

@article{Si2026CollaborativeRegistryPlanning,
  author  = {Si, Ziyou and Gu, Lin and Ju, Yunzhuo and Zeng, Deze and Jin, Hai},
  title   = {Collaborative multi-granularity distributed registry planning for fast container image pulling},
  journal = {Frontiers of Computer Science},
  year    = {2026},
  volume  = {20},
  number  = {10},
  pages   = {2010617},
  doi     = {10.1007/s11704-025-50350-y}
}

@article{Llama3.1,
  title={The {Llama} 3 Herd of Models},
  author={Dubey, Abhimanyu and Jauhri, Abhinav and Pandey, Abhinav and Kadian, Abhishek and Al-Dahle, Ahmad and Letman, Aiesha and Mathur, Akhil and Schelten, Alan and Yang, Amy and Fan, Angela and others},
  journal={arXiv preprint arXiv:2407.21783},
  year={2024}
}

@misc{lriplib-dataset,
  title  = {{L-RIPLIB}: An Industrial-Scale Resource Investment Problem Benchmark},
  author = {Hu, Yi-Xiang and Wang, Yuke and Wu, Feng and Huang, Zirui and Zeng, Shuli and Li, Xiang-Yang},
  howpublished = {doi: 10.57967/hf/9608},
  year   = {2026}
}

@inproceedings{hu2025ffcg,
  author    = {Yi-Xiang Hu and Feng Wu and Shaoang Li and
               Yifang Zhao and Xiang-Yang Li},
  title     = {{FFCG}: Effective and Fast Family Column Generation
               for Solving Large-Scale Linear Program},
  booktitle = {Proceedings of the AAAI Conference on Artificial Intelligence},
  volume    = {39},
  number    = {11},
  pages     = {11238--11245},
  year      = {2025}
}

@inproceedings{NEURIPS2025_31e89283,
 author = {Xie, Wantong and Hu, Yi-Xiang and Xu, Jieyang and Wu, Feng and Li, Xiangyang},
 booktitle = {Advances in Neural Information Processing Systems},
 pages = {34878--34905},
 publisher = {Curran Associates, Inc.},
 title = {{MURKA}: Multi-Reward Reinforcement Learning with Knowledge Alignment for Optimization Tasks},
 volume = {38},
 year = {2025}
}

\begin{biography}{yixianghu}
Yi-Xiang Hu received his B.E. degree in computer science and technology from the School of the Gifted Young, University of Science and Technology of China, China, in 2022. He is currently pursuing a Ph.D. degree
at the University of Science and Technology of China,
China. His research interests include optimization and data center network.
\end{biography}

\begin{biography}{yukewang}
Yuke Wang received his B.E. degree in computer science and technology from the School of Computer Science and Technology, Jilin University, China, in 2024. He is currently pursuing a Master's degree
at the University of Science and Technology of China, China. His research interests focus on optimization.
\end{biography}

\begin{biography}{fengwu}
Feng Wu received the B.E. and Ph.D. degrees from the University of Science and Technology of China (USTC), China, in 2006 and 2011, respectively. He is currently an Associate Professor with the School of Computer Science and Technology, USTC. His research interests include planning under uncertainty, multi-agent systems, reinforcement learning, and robotics.
\end{biography}

\begin{biography}{ziruihuang}
Zirui Huang received his B.E. degree in computer science and technology from the School of Computer Science and Technology, University of Science and Technology of China, China, in 2025. He is currently pursuing a Master's degree
at the University of Science and Technology of China,
China. His research interests  focus on optimization.
\end{biography}

\begin{biography}{shulizeng}
Shuli Zeng received his B.E. degree in computer science and technology from the School of the Gifted Young, University of Science and Technology of China, China, in 2023. He is currently pursuing a Ph.D. degree
at the University of Science and Technology of China,
China. His research interests include optimization and reinforcement learning.
\end{biography}

\begin{biography}{xiangyangli}
Xiang-Yang Li received his bachelor’s degree from the Department of Computer Science, Tsinghua University, in 1995, the bachelor’s degree from the Department of Business Management, Tsinghua University, in 1995, and his M.S. and Ph.D. degrees from the Department of Computer Science, University of Illinois at Urbana-Champaign, in 2000 and 2001, respectively. He is currently a Full Professor and the Executive Dean of the School of Computer Science and Technology, University of Science and Technology of China, Hefei, China. His research interests include the artificial intelligent Internet of Things, mobile computing, data sharing and trading, and privacy. He is a Fellow of IEEE. He is a Fellow of ACM. He is an ACM Distinguished Scientist.
\end{biography}

\end{document}